\documentclass[%
 reprint,
 amsmath,amssymb,
 aps,
]{revtex4-1}

\usepackage{graphicx}
\usepackage{dcolumn}
\usepackage{bm}

\usepackage[english]{babel}
\usepackage[utf8x]{inputenc}
\usepackage[T1]{fontenc}
\usepackage[colorinlistoftodos]{todonotes}
\usepackage{hyperref}
\usepackage{braket}
\usepackage{subcaption}
\usepackage{ragged2e}
\DeclareCaptionJustification{justified}{\justifying}
\usepackage{lipsum}
\DeclareMathAlphabet{\mathpzc}{OT1}{pzc}{m}{it}
\usepackage{mwe}
\usepackage{titlesec}
\usepackage{babel,blindtext}
\usepackage{mathrsfs}
\usepackage{gensymb}
\usepackage{sidecap}
\usepackage{float}

\begin{document}

\title{Noncyclic geometric phases and helicity transitions for neutrino oscillations in a magnetic field}

\author{Sandeep Joshi}
\email{sjoshi@barc.gov.in}
\author{Sudhir R. Jain}%
 \email{srjain@barc.gov.in}
\affiliation{%
 Nuclear Physics Division, Bhabha Atomic Research Centre, Mumbai 400085, India\\
Homi Bhabha National Institute, Anushakti Nagar, Mumbai 400094, India
}%

\begin{abstract}
We show that neutrino spin and spin-flavor transitions involve nonvanishing geometric phases. The geometric character of  neutrino spin rotation is explored by studying the neutrino spin trajectory in the projective Hilbert space representation and its relation to the geometric phase. Analytical expressions are derived for noncyclic geometric phases. Several calculations are performed for different cases of rotating and nonrotating magnetic fields in the context of solar neutrinos and neutrinos produced inside neutron stars. Also the effects of adiabaticity, critical magnetic fields and cross boundary effects in the case of neutrinos emanating out of neutron stars are examined. 
\end{abstract}	

\maketitle

\section{Introduction}
The study of neutrino oscillations in vacuum, matter and magnetic fields is a widely discussed topic in high energy physics. The experimental confirmation of 
the neutrino oscillation hypothesis \cite{Fukuda:1998mi,*Ahmad:2001an,*Ahmad:2002jz} proves that neutrinos are massive particles. Since the standard model (SM) assumes neutrinos to be massless, in the simplest extension of the SM, right-handed components of the neutrino fields are introduced to account for the mass of the neutrinos. In this minimally extended SM, neutrinos acquire nonzero magnetic dipole moments due to coupling with the photon at one-loop level \cite{Marciano:1977wx, Lee:1977tib, Kayser:1982br}. This results in spin and spin-flavor oscillations between left- and right-handed neutrino states in the presence of sufficiently strong magnetic fields. This possibility has been discussed for the cases of both Dirac \cite{Fujikawa:1980yx} and Majorana neutrinos \cite{Schechter:1981hw}. Since the right-handed Dirac neutrino states are singlets under the $SU(2)_L$ gauge group, they do not participate in weak interactions and hence are sterile and do not show up in the experiments. For the case of Majorana neutrinos, even though diagonal magnetic moments vanish \cite{Schechter:1981hw}, there could be nonzero off-diagonal magnetic moments which result in transitions between neutrinos of different flavors and helicities, commonly termed as $\nu \leftrightarrow \bar{\nu}$ transitions. The spin and spin-flavor oscillations may also lead to resonant conversion mechanisms, similar to the Mikheyev-Smirnov-Wolfenstein effect \cite{Mikheev:1986if}, the possibility of which has been extensively discussed in various scenarios including solar \cite{Lim:1987tk,Akhmedov:1988uk,Akhmedov:1988hd,Pulido:1989ym,Balantekin:1990jg,Akhmedov:1991vj,Pal:1991pm,Akhmedov:1993sh,Guzzo:1993np,Miranda:2000bi,Barranco:2002te} and supernova neutrinos \cite{Voloshin:1988xu,Likhachev:1990ki,Akhmedov:1996ec,Ando:2003is,Ahriche:2003wt,Akhmedov:2003fu}(see \cite{Giunti:2014ixa} for a detailed list of references).

Associated with the phenomenon of neutrino oscillations in vacuum, matter, and magnetic fields is the emergence of geometric phase, which has been explored by many authors in various settings     \cite{Nakagawa:1987ys,Vidal:1990fr,Aneziris:1990my,Smirnov:1991ia,Naumov:1991ju,Guzzo:1992ij,Naumov:1993vz,Blasone:1999tq,  Wang:2000ep,He:2004zc,Law:2007fb,Mehta:2009ea,Dajka:2011zz,Joshi:2016unj,Capolupo:2016idi,Johns:2016wjd,Wang:2015tqp}. The concept of geometric phase emerges from the idea that the phase factor acquired by the wavefunction of a system undergoing quantum evolution has a part that is dynamical and a part that is path dependent or geometric in nature. Berry in his seminal paper \cite{Berry:1984jv} showed that for systems undergoing cyclic, adiabatic evolution this path-dependent geometric phase can have observable consequences. The definition of geometric phase has been subsequently generalized to include nonadiabatic, noncyclic and nonunitary evolutions \cite{Aharonov:1987gg,Samuel:1988zz,aitchison1992real,pati1995new}. 

The phenomena of geometric phase emerges in  a wide range of classical and quantum systems and dates back to the work of Pancharatnam \cite{pancharatnam1956generalized} who demonstrated that when a polarized light is passed through a series of polarizers such that initial polarization is finally restored, the final polarized state acquires an additional phase. This additional phase is equal to half the solid angle subtended by the curve representing the polarization states on a Poincaré sphere.  The geometric phase since then has been observed in a wide range of systems such as molecular physics \cite{Mead:1992zz}, neutron spin rotation \cite{Bitter:1987zz}, photon propagation in helically wound optical fiber \cite{Tomita:1986zz}, and response functions of many-body Fermionic systems \cite{Jain:1998zz}. In particle physics, in addition to neutrinos,  the importance of geometric phase has been explored in the context of supersymmetric quantum mechanics \cite{Pedder:2007ff}, CPT (Charge conjugation, Parity and Time reversal) violation in meson systems \cite{Capolupo:2011rd}, and  axion-photon mixing \cite{Capolupo:2015cga}.

To briefly describe the notion of geometric phase, we consider the parallel transport of a vector on a curved surface. If the vector is transported along a closed curve then it gives rise to holonomy due to change in the vector's orientation. This holonomy arises due to the curvature of the surface and depends only on the area enclosed by the curve and the geometry of the curved surface. This idea of parallel transport and holonomy associated with the curve can be extended to quantum regime. Let $\mathcal{H}$ denote the set of possible states of a quantum system, known as Hilbert space. Since two vectors $\Ket{\psi}, e^{i \phi} \Ket{\psi} \in \mathcal{H}$ differing only by the phase factor correspond to the same physical state, we define a projection map $\pi: \mathcal{H} \rightarrow \mathcal{P}$ defined by $\pi(e^{i \phi}\Ket{\psi}) = \pi(\Ket{\psi})\in \mathcal{P}$ $ \forall$ $ \phi \in \mathbb{R}$ and $\Ket{\psi} \in \mathcal{H}$. Thus vectors differing by a phase factor correspond to the same element in projective Hilbert space $\mathcal{P}$. Thus the evolution of the state vector $\Ket{\psi(s)} \in \mathcal{H}$, where $s$ is a parameter, such that $\Ket{\psi(s)} = e^{i \phi} \Ket{\psi(0)}$ describes a closed  curve  $\mathcal{C}$ in  $\mathcal{P}$. However, since we are not interested in dynamical phases, we define vector $\Ket{\xi(s)}$ which is basically $\Ket{\psi(s)}$ with its dynamical phase removed. The parallel transport of $\Ket{\xi(s)}$ can be obtained by demanding that the magnitude of $\Ket{\xi(s)}$ is preserved and that $\Ket{\xi(s)}$ and $\Ket{\xi(s+ ds)}$ have the same phase. These two conditions give  
\begin{equation}\label{connection}
\Im \Braket{\xi(s)|\frac{d}{ds}|\xi(s)} = 0, 
\end{equation}
where $\Im$ denotes the imaginary part. Eq.\eqref{connection} defines a connection over $\mathcal{P}$ and implies that the parallel transport forbids local rotations along the curve. The curve described by $\Ket{\xi(s)}$ in $\mathcal{H}$ is called the horizontal lift of $\mathcal{C}$ in $\mathcal{P}$. The horizontal lift of closed curve in $\mathcal{C}$ may be open in $\mathcal{H}$ such that $\Ket{\xi(s)} = e^{i \beta} \Ket{\xi(0)}$. Here, the factor $\beta$ arises from the holonomy of the connection and is the geometric phase associated with $\mathcal{C}$.

The geometric phase depends only on the geometry of the curve in projective Hilbert space and for the case of cyclic evolution it can be interpreted in terms of surface area enclosed by the closed curve $\mathcal{C}$ in $\mathcal{P}$. For the more general case of noncyclic evolution the geometric phase is proportional to the surface area enclosed by the curve in $\mathcal{P}$ which is composed of two parts: the open curve $\mathcal{C} = \{\pi(\Ket{\xi(s)}) \in \mathcal{P} | s \in [s_1, s_2] \subset \mathbb{R}\}$ describing the evolution from the initial point $\Ket{\xi(s_1)}$ to the final point $\Ket{\xi(s_2)}$, and a geodesic curve in $\mathcal{P}$ joining  $\pi(\Ket{\xi(s_2)})$ to $\pi(\Ket{\xi(s_1)})$. 

Even though the geodesic closure approach, as formulated in \cite{Samuel:1988zz}, gives an elegant and robust definition of geometric phase, the calculations of geodesic can be tedious and in some cases may lead to inconsistent results \cite{GarciadePolavieja:1998vy}. An equivalent approach to calculate the geometric phase for any general nonadiabatic, noncyclic evolution has been developed by Mukunda and Simon \citep{Mukunda:1991rc}. Their treatment is based entirely on the kinematics, and the geometric phase is defined as the property of curves in Hilbert space. If $\mathscr{C}$ is any one-parameter smooth curve of unit vectors $\Ket{\psi(s)} \in \mathcal{H}$, where $s \in[s_1, s_2]\subset \mathbb{R}$, then the geometric phase associated with the corresponding curve $\mathcal{C} \in \mathcal{P}$ is defined by the functional
\begin{equation} \label{gp1}
\phi_g[\mathcal{C}] = \arg\braket{\psi(s_1)|\psi(s_2)}- \Im \int_{s_1}^{s_2} ds \braket{\psi(s)|\dot{\psi}(s)},
\end{equation}
where $\Ket{\dot{\psi}(s)}$ denotes the derivative with respect to s. The geometric phase $\phi_g[\mathcal{C}]$ can easily be shown to be gauge and reparametrization invariant. The two terms on the right-hand side of Eq.\eqref{gp1} are, respectively the total and dynamical phase associated with the curve $\mathscr{C}$, and the difference between the two gives the geometric phase along $\mathcal{C}$.

In the present work we analyze the noncyclic geometric phases that arise due to neutrino oscillations in magnetic fields and matter. In particular, we first perform explicit calculations for the geometric phases that arise due to spin and spin-flavor precession of neutrinos propagating in a medium with constant density and uniformly twisting magnetic fields. We then study the case of geometric phase acquired by neutrinos produced inside and emanating out of a neutron star, with realistic density and magnetic field profiles. We also study the condition of adiabaticity, the effects of magnetic field rotation and cross boundary effects on geometric phases and neutrino helicity transitions. 
This is a generalization of our previous work \cite{Joshi:2016unj}, where we calculated the cyclic Berry phase for the neutrino propagation in a magnetic field rotating arbitrarily about the direction of motion of the neutrino. There it was shown that as the rotating magnetic field traces a closed curve in the parameter space, the neutrino eigenstates can develop a significant geometric phase if the magnetic field is sufficiently strong ($ \sim 10^7$ G or more). Such magnetic fields are usually encountered by neutrinos propagating though astrophysical environments such as core-collapse supernova, neutron stars, and gamma ray bursts.

The calculations pertaining to spin and spin-flavor oscillations of neutrinos depend on the  strength of one-loop magnetic dipole moments. In the minimally extended SM, the diagonal magnetic moments of the Dirac neutrinos to one-loop order have been calculated to be \cite{Marciano:1977wx,Lee:1977tib} $\mu_\nu= 3.2 \times 10^{-19} (m_\nu/eV) \mu_B $, where $m_\nu$ is the neutrino mass. However, this value is much smaller than the sensitivity of present experiments. The current best experimental constraint on the neutrino magnetic moment comes from the GEMMA experiment which puts an upper bound of $\mu_\nu< 2.9 \times 10^{-11} \mu_B$ \cite{Beda:2012zz,*Beda:2013mta}. A recent analysis \cite{Canas:2015yoa} of the Borexino data obtains the bound on the Majorana transition  magnetic moment at $\mu_\nu \leq 3.1 \times 10^{-11} \mu_B$. Also there are various solar, reactor, and accelerator neutrino experiments that obtain different bounds on neutrino magnetic  moments \cite{Giunti:2015gga}. On the theoretical side, various models have been proposed that derive the bounds on the neutrino magnetic moment as large as $10^{-10} \mu_B$ (see \cite{Giunti:2014ixa} for a detailed review). Here we take the value of the neutrino magnetic moment, $\mu_\nu = 10^{-11} \mu_B$.   
 
\section{Neutrino spin and spin-flavor evolution}
In the quasiclassical approach, neutrino spin evolution in an electromagnetic field is described by the generalized Bargmann-Michel-Telegdi equation \cite{Bargmann:1959gz}. For the system $\Ket{\nu}= (\nu_R, \nu_L)^T$ with two helicity components of neutrinos propagating in the presence of magnetic field $\vec{B}$ in matter, the effective Hamiltonian is given by \cite{Egorov:1999ah}
\begin{equation} \label{hamiltonian}
H=  (\vec{\sigma}.\vec{n})\Big(\frac{\Delta m^2 A}{4E}- \frac{\Delta V}{2}\Big)- \mu \vec{\sigma}.\Big(\vec{B}- \big(1+ \frac{1}{\gamma}\big)(\vec{B}.\vec{n})\vec{n}\Big),
\end{equation}
where $\vec{n}$ is the direction of propagation of the neutrino, $\vec{\sigma}$ are Pauli spin matrices, $\Delta V= V_L- V_R$ ($V_L, V_R$ being potentials due to coherent forward scattering of the neutrinos off matter particles \cite{Wolfenstein:1977ue} for left- and right-handed neutrinos respectively), $\Delta m^2= m_R^2- m_L^2$, $A$ is a function of 
neutrino mixing angle $\theta$, and $E$ is the neutrino energy. In Eq.\eqref{hamiltonian} the terms proportional to identity matrix are omitted.

Assuming the neutrinos to be propagating along the z direction, the evolution of the state $\Ket{\nu}$  can be described by the Schr\"{o}dinger-like  equation \cite{Okun:1986na} 
\begin{equation}\label{evol1}
i\frac{\partial \Ket{\nu(t)}}{\partial t}= H(t) \Ket{\nu(t)}.
\end{equation}
Since the longitudinal component of the magnetic field in Eq.\eqref{hamiltonian} is suppressed by a factor of $1/\gamma$, for relativistic neutrinos this term is negligible and we consider the magnetic field rotating clockwise about the neutrino direction in the transverse plane $B_\perp= Be^{i\phi}$. The evolution equation \eqref{evol1} can now be rewritten as 
\begin{equation} \label{evol2}
i\frac{\partial}{\partial z}\begin{pmatrix}
 \nu_R \\ \nu_L
\end{pmatrix}= - \begin{pmatrix}
V(z)/2 && \mu B(z)e^{-i\phi(z)}\\
\mu B(z)e^{i\phi(z)} && -V(z)/2
\end{pmatrix}\begin{pmatrix}
\nu_R \\ \nu_L
\end{pmatrix},
\end{equation}
where \begin{equation}
V= \Delta V- \frac{\Delta m^2 A}{2 E},
\end{equation} 
and the distance $z$ along the neutrino trajectory is approximated with time $t$. 

Transforming to the rotating frame of the field, and using 
\begin{equation}\label{trans1}
\Ket{\nu}= U \Ket{\psi}= \exp (-i\sigma_3 \phi/2) \Ket{\psi},
\end{equation}
we get an evolution equation in the rotating frame,
\begin{equation}\label{evol3}
\begin{aligned}[b]
i \frac{\partial \Ket{\psi}}{\partial z}=& (U^{-1}HU- iU^{-1}\frac{dU}{dz})\Ket{\psi}\\ =& -\frac{1}{2}\big[(V+\dot{\phi})\sigma_3+ (2\mu B) \sigma_1\big] \Ket{\psi},
\end{aligned}
\end{equation}
where $\dot{\phi}= d\phi/dz$. For the case of neutrino propagation in matter with constant density and in a magnetic field of constant strength and uniform twist, i.e., constant $V$, $B$, and $\dot{\phi}$, Eq.\eqref{evol3} can be integrated analytically and we obtain
\begin{equation}
\begin{pmatrix}
 \psi_R(z) \\ \psi_L(z)
\end{pmatrix}= \exp\bigg[\frac{i}{2}\bigg((V+\dot{\phi})\sigma_3+ 2\mu B \sigma_1\bigg)z\bigg]
\begin{pmatrix}
\psi_R(0) \\ \psi_L(0)
\end{pmatrix}.
\end{equation}
Using properties of Pauli matrices this can be written as
\begin{equation}
\begin{aligned}[b]
\begin{pmatrix}
 \psi_R(z) \\ \psi_L(z)
\end{pmatrix}=& \bigg[ \cos\bigg(\frac{\delta E_m z}{2}\bigg)+ \frac{i}{\delta E_m}\bigg((V+\dot{\phi})\sigma_3+ 2\mu B \sigma_1 \bigg)\\& \sin\bigg(\frac{\delta E_m z}{2}\bigg) \bigg] \begin{pmatrix}
\psi_R(0) \\ \psi_L(0)
\end{pmatrix},
\end{aligned}
\end{equation}
where \begin{equation}
\delta E_m= \sqrt{(V+\dot{\phi})^2+ (2\mu B)^2}
\end{equation}
gives the energy splitting of the eigenstates. If a neutrino is initially created in the left-helicity state, i.e., $\Ket{\nu(0)}= (0 \quad  1)^T$, then after traveling a distance $z$ in the magnetic field,  the neutrino eigenstate will be an admixture of left- and right-handed components:
\begin{equation} \label{state_nuL}
\Ket{\nu(z)}= \begin{pmatrix} i e^{-i\phi(z)/2}\sin 2\theta_m \sin \left(\frac{\delta E_m z}{2}\right) \\ \\ e^{i\phi(z)/2}\big( \left( \frac{\delta E_m z}{2}\right) - i \cos 2\theta_m \sin \left(\frac{\delta E_m z}{2}\right)\big)    \end{pmatrix}.
\end{equation}
Here, we have taken the reference direction as $\phi(0)= 0$ and $\theta_m$ denotes the mixing angle between $\psi_R$ and $\psi_L$,
\begin{equation}\label{mixing}
\tan 2\theta_m= \frac{2\mu B}{V+\dot{\phi}}\hspace{1mm}.
\end{equation}
If a beam of left-handed neutrinos starts at $z=0$, the transition probability at a distance $z$ is given by
\begin{equation}\label{prob1}
P(\nu_L\rightarrow\nu_R; z)= |\nu_R(z)|^2=\sin^2 2\theta_m \sin^2	\bigg(\frac{\delta E_m	z}{2}\bigg).
\end{equation}
Thus neutrino propagation in magnetic fields results in an oscillation in the $\nu_L-\nu_R$ basis with a length scale of $2 \pi / \delta E_m$.
For $\theta_m= \pi/4$ the mixing is maximal and the amplitude of the transition probability becomes unity. Eq.\eqref{mixing} gives the condition for resonant $\nu_R \leftrightarrow \nu_L$ conversion 
\begin{align}
V+\dot{\phi}= 0,\label{resonance}\\
\mbox{or}\qquad  \Delta V- \frac{\Delta m^2}{2E}A + \dot{\phi}= 0.
\end{align}
The effects of the variation of the twisting field on the transition probability has been explored in detail in \cite{Akhmedov:1991vj}.
If we now include the effects of the three neutrino flavor, the effective Hamiltonian becomes a $6 \times 6$ matrix that can be written as
\begin{equation}
H = H_0+ H_{wk}+ H_B,
\end{equation}
where $H_0$ is the vacuum term that is the same for both Dirac and Majorana neutrinos, given by
\begin{equation}
H_0 = \frac{1}{2E} \begin{pmatrix}
U & 0\\
0 & U
\end{pmatrix}\begin{pmatrix}
M^2 & 0\\
0 & M^2
\end{pmatrix}\begin{pmatrix}
U^{\dagger} & 0\\
0 & U^{\dagger}
\end{pmatrix},
\end{equation}
where $M^2$ is the mass matrix, $M^2= \mbox{Diag} (0, \Delta  m_{21}^2, \Delta m_{31}^2)$, and $U$ is the unitary mixing matrix that can be suitably parameterized. $H_{wk}$ is the matter potential term given by
\begin{equation}
\begin{aligned}[b]
H_{wk}= \alpha \rho \begin{cases} \mbox{Diag}\big(Y_e, 0, 0, (1-Y_e)/2,  (1-Y_e)/2,\\ (1-Y_e)/2\big) \quad \mbox{for Dirac neutrinos}\\
 \mbox{Diag}\big(Y_e, 0, 0, 1-2Y_e,  1-Y_e, 1-Y_e\big) \\ \mbox{for Majorana neutrinos}
\end{cases},
\end{aligned}
\end{equation}
where $Y_e$ is the electron fraction, $\rho$ is the density of the medium, and $\alpha= \sqrt[]{2} G_F/m_N$ is constant 
Finally, $H_B$ denotes the neutrino coupling with magnetic field
\begin{equation}
H_B= \begin{pmatrix}
0 & M_\mu ^{\dagger} Be^{i\phi}\\
\pm M_\mu Be^{-i\phi} & 0
\end{pmatrix},
\end{equation}
where the $\pm$ sign refers to Dirac and Majorana neutrinos, respectively and $M_\mu$ is the magnetic moment matrix given by
\begin{equation}
M_\mu= \begin{cases} \begin{pmatrix}
\mu_{ee} & \mu_{e\mu} & \mu_{e \tau} \\
\mu_{e\mu} & \mu_{\mu\mu} & \mu_{\mu \tau} \\
\mu_{e\tau} & \mu_{\mu\tau} & \mu_{\tau \tau} 
\end{pmatrix} \quad \mbox{for Dirac neutrinos}\\ \\ 
\vspace{1mm}
 \begin{pmatrix}
0 & \mu_{e\mu} & \mu_{e \tau} \\
-\mu_{e\mu} & 0 & \mu_{\mu \tau} \\
-\mu_{e\tau} & -\mu_{\mu\tau} & 0	
\end{pmatrix} \quad \mbox{for Majorana neutrinos}
\end{cases}.
\end{equation}
If a neutrino is produced as a left-handed electron neutrino and propagates in a magnetic field, a combination of diagonal and off-diagonal magnetic moments leads to the following spin and spin-flavor transitions between different neutrino states: \\ 
\noindent 
{\it Dirac neutrinos}: $\nu{_e{_L}} \rightarrow \nu{_e{_R}},\nu{_e{_L}} \rightarrow \nu{_\mu{_R}}, \nu{_e{_L}} \rightarrow\nu{_\tau{_R}}$,\\
{\it Majorana neutrinos} : $\nu{_e{_L}} \rightarrow \bar{\nu}_\mu,\nu{_e{_L}} \rightarrow \bar{\nu}_\tau $.

Moreover, the condition for resonant transitions Eq.\eqref{resonance} can be written as
\begin{equation}
\alpha \rho Y_e^{\rm eff} - \frac{\Delta m^2 A}{2E}+
\dot{\phi}= 0 .
\end{equation}
where \begin{equation}
Y_e^{\rm eff}= \begin{cases}
(3 Y_e-1)/2 \quad \mbox{for}\quad \nu{_e{_L}}\leftrightarrow \nu_{e{_R},\mu{_R},\tau{_R}},\\
(2 Y_e-1) \quad \mbox{for} \quad  \nu_{e{_L}}\leftrightarrow \bar{\nu}_{\mu,\tau}.
\end{cases}
\end{equation}
We now examine the geometric phases associated with the spin and spin-flavor evolution of the neutrinos propagating in magnetic fields and matter. 

\section{Neutrino spin rotation on the Bloch sphere and noncyclic geometric phases}

The dynamics of the neutrino spin rotation in a magnetic field can be described by vector $\textbf{n}= \Braket{\nu|\bm{\sigma}|\nu}$. In the two component formalism the equation describing the dynamics of $\textbf{n}$ is equivalent to a Schr\"{o}dinger-like  equation \eqref{evol2}, and is given by \cite{Feynman:1957zz}
\begin{equation} \label{torque}
\frac{d \textbf{n}}{dz}= \textbf{n} \times \textbf{B}_\textbf{eff},
\end{equation}
where $\textbf{B}_\textbf{eff}= \sqrt[]{V^2+ (2 \mu B)^2}(\sin \chi \cos \phi, \sin \chi \sin \phi, \cos \chi)$; $\chi= \tan^{-1}(2 \mu B/V)$. The path of the effective magnetic field $\textbf{B}_\textbf{eff}$ describes a circle around the z axis in the parameter space, which is the unit 2-sphere $S^2$. For the case of a medium with a uniformly twisting magnetic field and constant density as discussed in the previous section, Eq.\eqref{torque} can be solved analytically, and the resulting solution \textbf{n(z)} can be plotted in the Bloch sphere representation. In this representation, any neutrino state $\Ket{\nu}$ corresponds to a point on $S^2$ and is determined by the unit vector $\textbf{n}$. The orthogonal states $\Ket{\nu_L}$ and $\Ket{\nu_R}$ correspond to  two antipodal points on $S^2$.   

To solve Eq.\eqref{torque}, we define a vector $\textbf{n}_\textbf{R}= \textbf{n}\cdot \textbf{R}$, where $\textbf{R}$ is the rotation matrix  
\begin{equation} 
\textbf{R}= \begin{pmatrix}
\cos \phi & \sin \phi & 0\\
-\sin \phi & \cos \phi & 0\\
0 & 0 & 1
\end{pmatrix}.
\end{equation}
Substituting this in Eq.\eqref{torque} we obtain a time-independent differential equation for $\textbf{n}_\textbf{R}$, which can be integrated to give the solution for Eq.\eqref{torque} as the superposition of two rotations, 
\begin{equation} \label{torque_2}
\textbf{n}(z)= \textbf{R}^T \cdot \textbf{R}_1 \cdot \textbf{n}(0) ,
\end{equation}  
where $\textbf{R}_1$ is given by
\begin{widetext}
\[
\textbf{R}_1= \begin{pmatrix}
\sin^2 2 \theta_m+ \cos^2 2 \theta_m \cos \phi_p & \cos 2 \theta_m \sin \phi_p & \frac{1}{2}\sin 2 \theta_m(1- \cos \phi_p)\\
- \cos 2 \theta_m \sin \phi_p & \cos \phi_p & \sin 2 \theta_m \sin \phi_p\\
\frac{1}{2}\sin 2 \theta_m(1- \cos \phi_p)& -\sin 2 \theta_m \sin \phi_p& \cos^2 2 \theta_m + \sin^2 2 \theta_m \cos \phi_p
\end{pmatrix},
\]
\end{widetext}
where we have defined the precession phase as $\phi_p = \delta E_m z$, such that $\phi_p$ varies from $0$ to $2 \pi$ during one oscillation length. 

\begin{figure*}[ht]
\begin{subfigure}[b]{0.475\textwidth} \label{fig:(a)}
            \centering
            \includegraphics[scale=0.5]{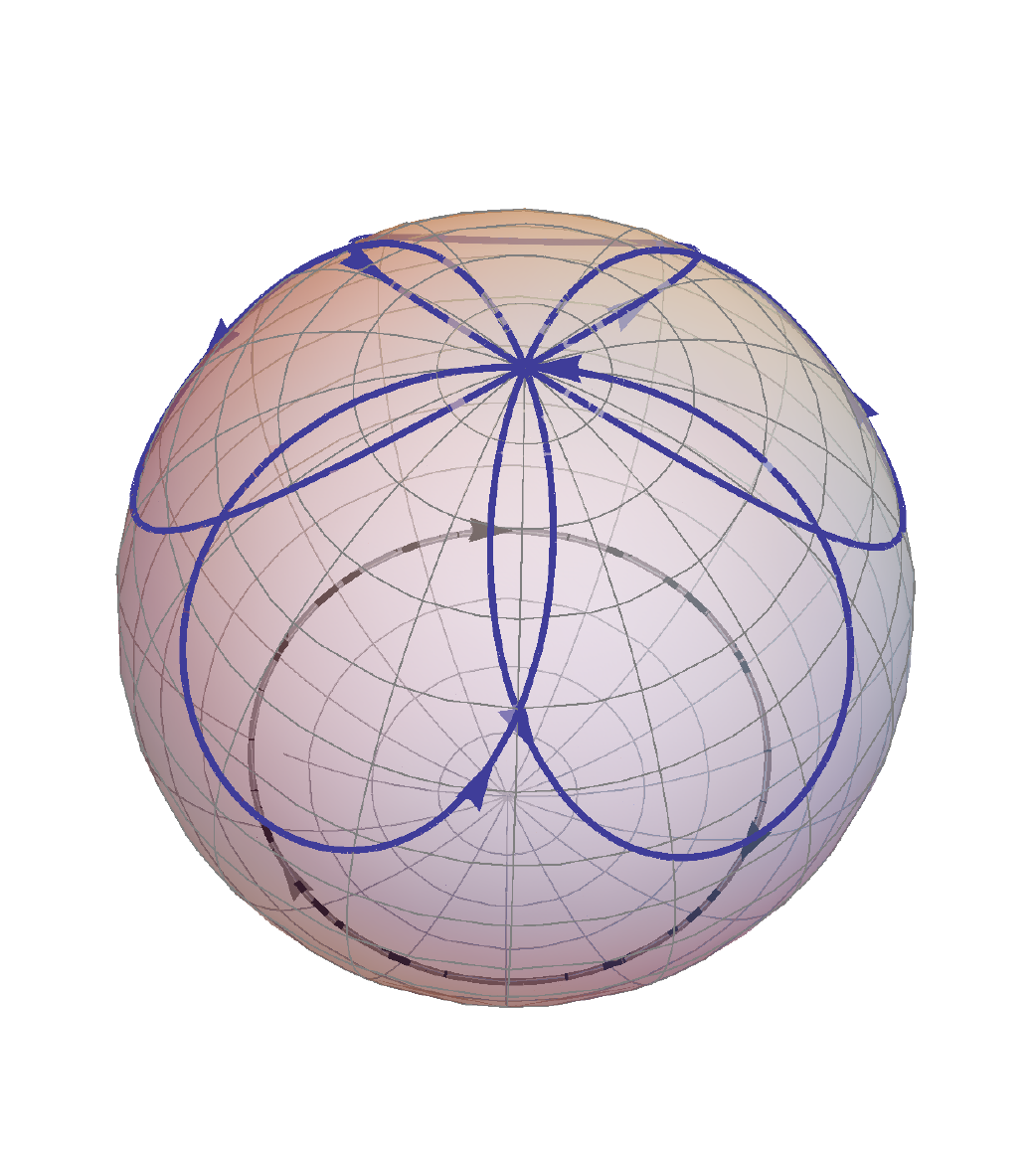}
            \caption[]
            {\centering{\small $\dot{\phi_p}= 5 \dot{\phi}$}}       
        \end{subfigure}
        \hfill
        \begin{subfigure}[b]{0.475\textwidth}   \label{fig:(b)}
            \centering 
            \includegraphics[scale=0.5]{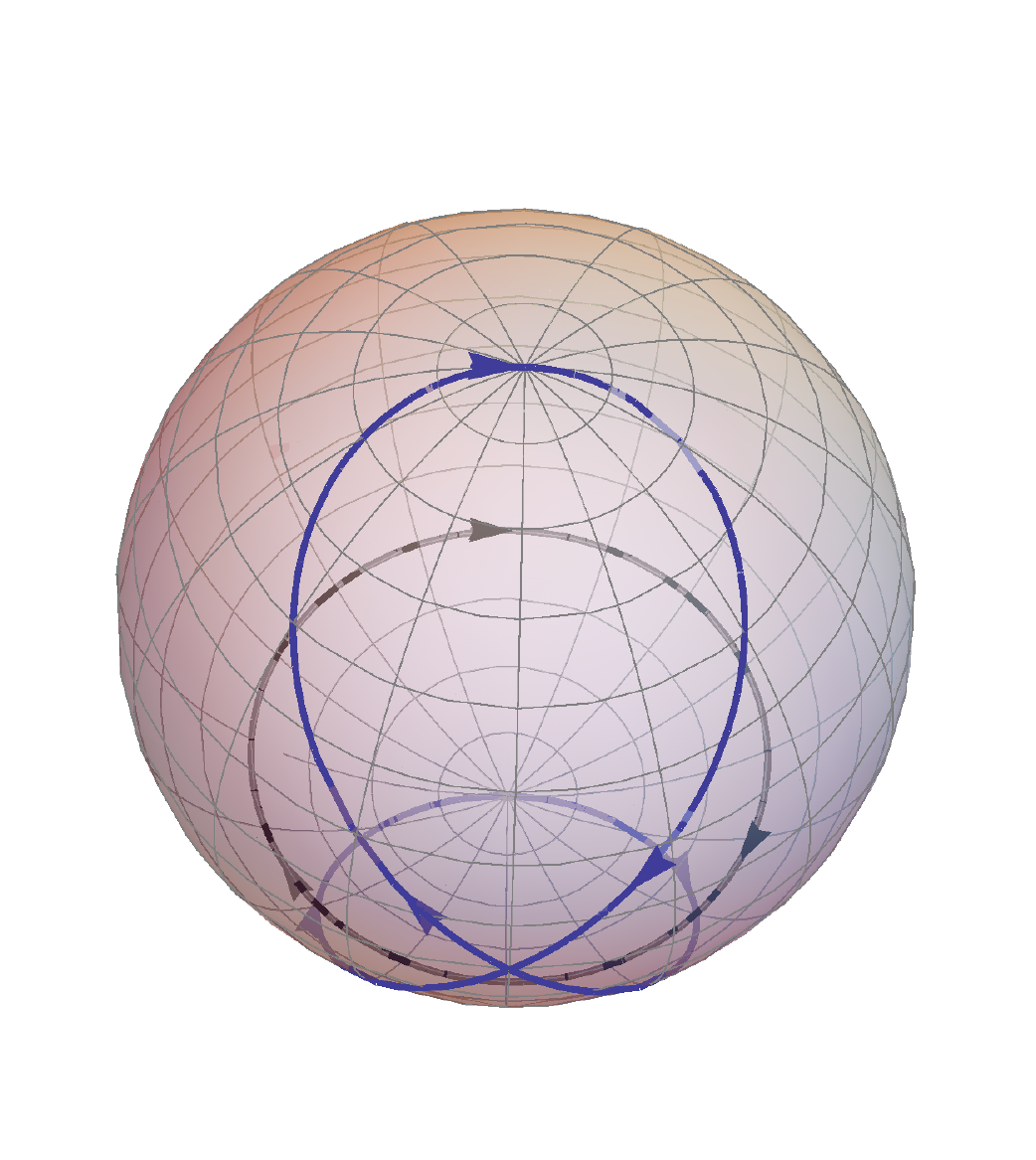}
            \caption[]%
            {\centering{\small $V= - \dot{\phi}$}}    
        \end{subfigure}
        \vskip\baselineskip
        \begin{subfigure}[b]{0.475\textwidth}    \label{fig:(c)}
            \centering 
            \includegraphics[scale=0.5]{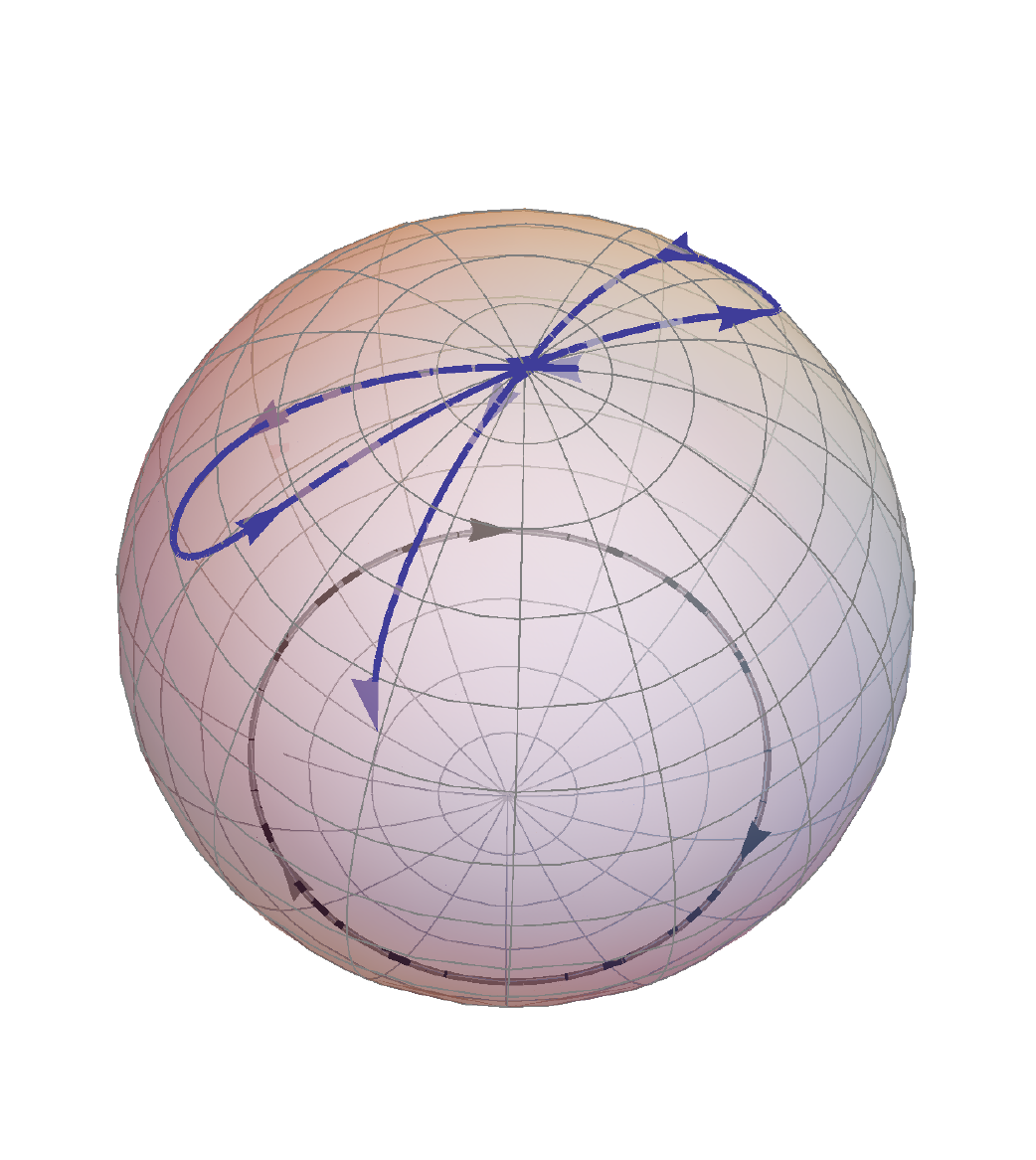}
            \caption[]%
            {\centering{\small $\dot{\phi} = 100$}}    
        \end{subfigure}
        \quad
        \begin{subfigure}[b]{0.475\textwidth}    \label{fig:(d)}
            \centering 
            \includegraphics[scale=0.5]{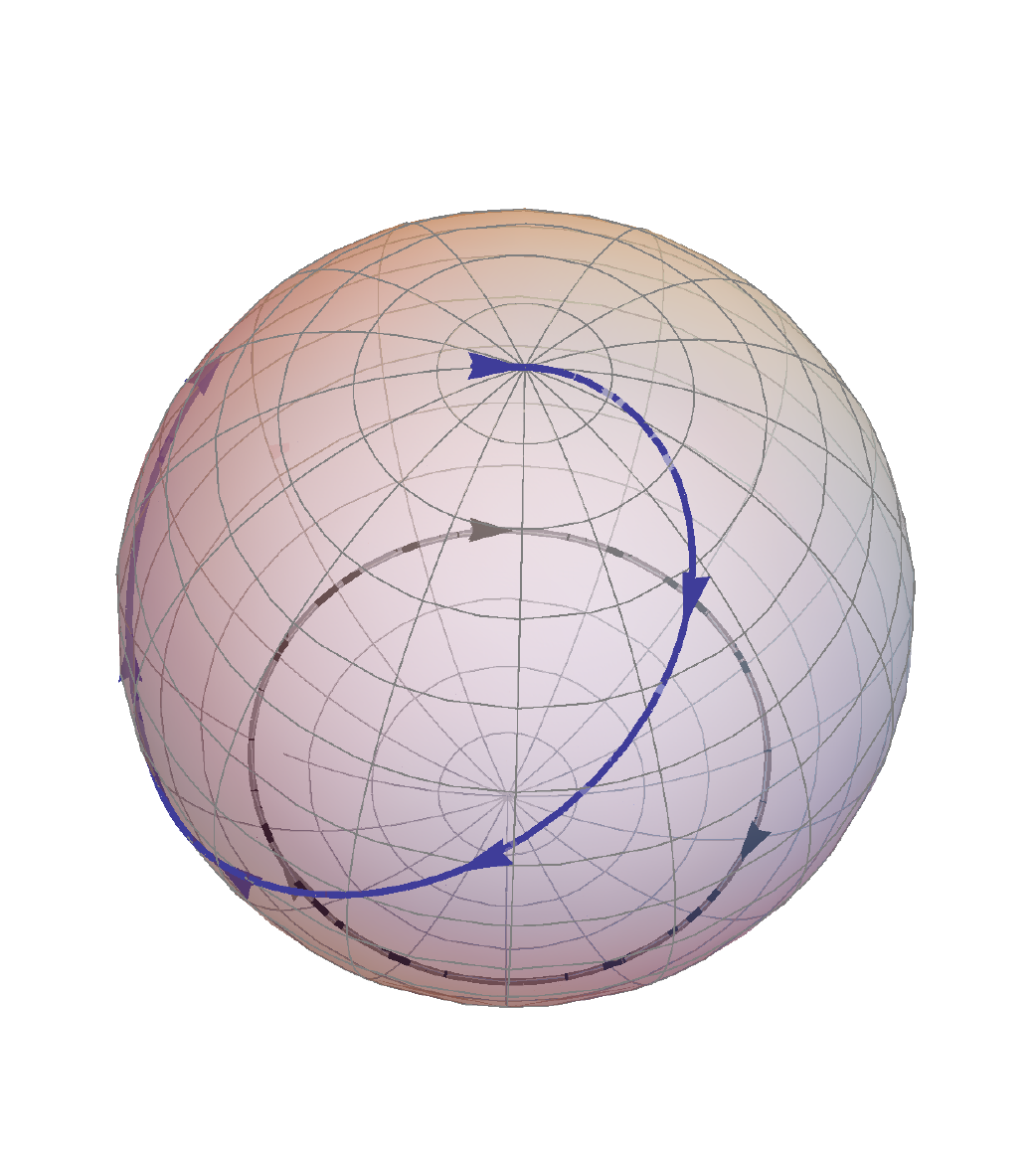}
            \caption[]%
            {\centering{\small $\dot{\phi} = -200$}}    
        \end{subfigure}  
        \caption{ Bloch sphere representation of neutrino spin rotation. Initially the neutrinos are produced in the left helicity state which corresponds to a point on the pole of the sphere. Under the effect of matter and magnetic field, neutrinos undergo spin-precession $\nu_{eL} \rightarrow \nu_{eR}$ and neutrino spin-vector $\textbf{n}$ traces out cyclic [(a) and (b)] and noncyclic curves [(c) and (d)] on the Bloch sphere depending on the relative values of $\dot{\phi}_p$ and the parameters of $\textbf{B}_\textbf{eff}$. The circular curve describes the path of $\textbf{B}_\textbf{eff}$. The rotation frequency is in units of $\pi$/R, and the positive and negative signs of $\dot{\phi}$ correspond to clockwise and anticlockwise rotation of the magnetic field about the neutrino direction respectively. We used the following parameters: electron number density $n_e= 10^{24}{\rm g/cm}^3$, neutron number density $n_n = n_e/6$, matter potential $V= \sqrt[]{2} G_F (n_e- n_n/2)$, and magnetic field strength $B = 10^6$ G.}
 \label{fig:bloch}
 \end{figure*}
The matrix $\textbf{R}_1$ represents a precession about  the direction of $\textbf{B}_\textbf{eff}$ at an angle $2\theta_m$ and at a rate $\dot{\phi}_p$, and  $\textbf{R}$ represents a precession about the direction of  propagation of the neutrino at a rate, $\dot{\phi}$. These two precessions combine to give the evolution of the spin-vector $\textbf{n}$, which may be plotted on the Bloch sphere. The curve traced by the spin-vector $\textbf{n}$ on the Bloch sphere, as the magnetic field rotates by $2 \pi$, is noncyclic in general. However, for the special case when the two precession rates $\dot{\phi}_p$ and $\dot{\phi}$ are commensurable i.e. $\dot{\phi}_p = k \dot{\phi}$ for some $k \in \mathbb{Q}$, the evolution becomes cyclic. Different cases for cyclic and noncyclic evolution are shown in Fig. \ref{fig:bloch} for the case of spin precession $\nu_{eL} \rightarrow \nu_{eR}$ of left-handed electron neutrinos produced in the Sun and propagating outwards under the influence of matter and magnetic fields. As a first order calculation we assume a constant density and magnetic field profile for the Sun and parametrize the rotation frequency of the magnetic field as $\dot{\phi} = \pi/ f R$, where $R$ is the radius of the Sun.

\begin{figure*}[ht]
\captionsetup{justification=raggedright,singlelinecheck=false}
\begin{subfigure}{.5\textwidth}
  \includegraphics[width=0.9\linewidth]{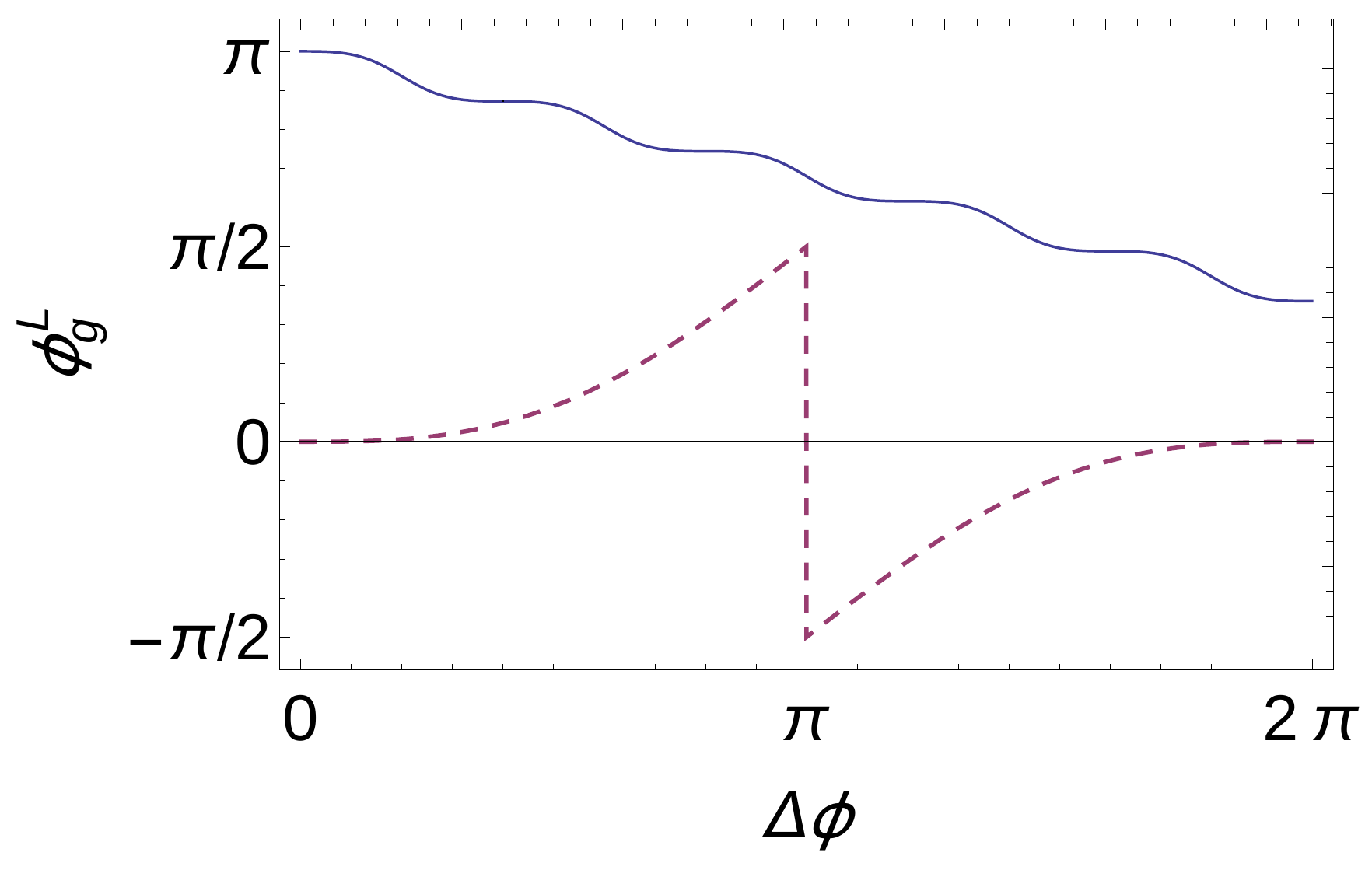}
  \caption{Cyclic geometric phase: Solid curve corresponds to the cyclic case $\dot{\phi}_p = 5 \dot{\phi}$ and the dotted curve corresponds to the resonant condition $V= - \dot{\phi}$}
  \label{fig:gp(a)}
\end{subfigure}%
\begin{subfigure}{.5\textwidth}
  \includegraphics[width=.9\linewidth]{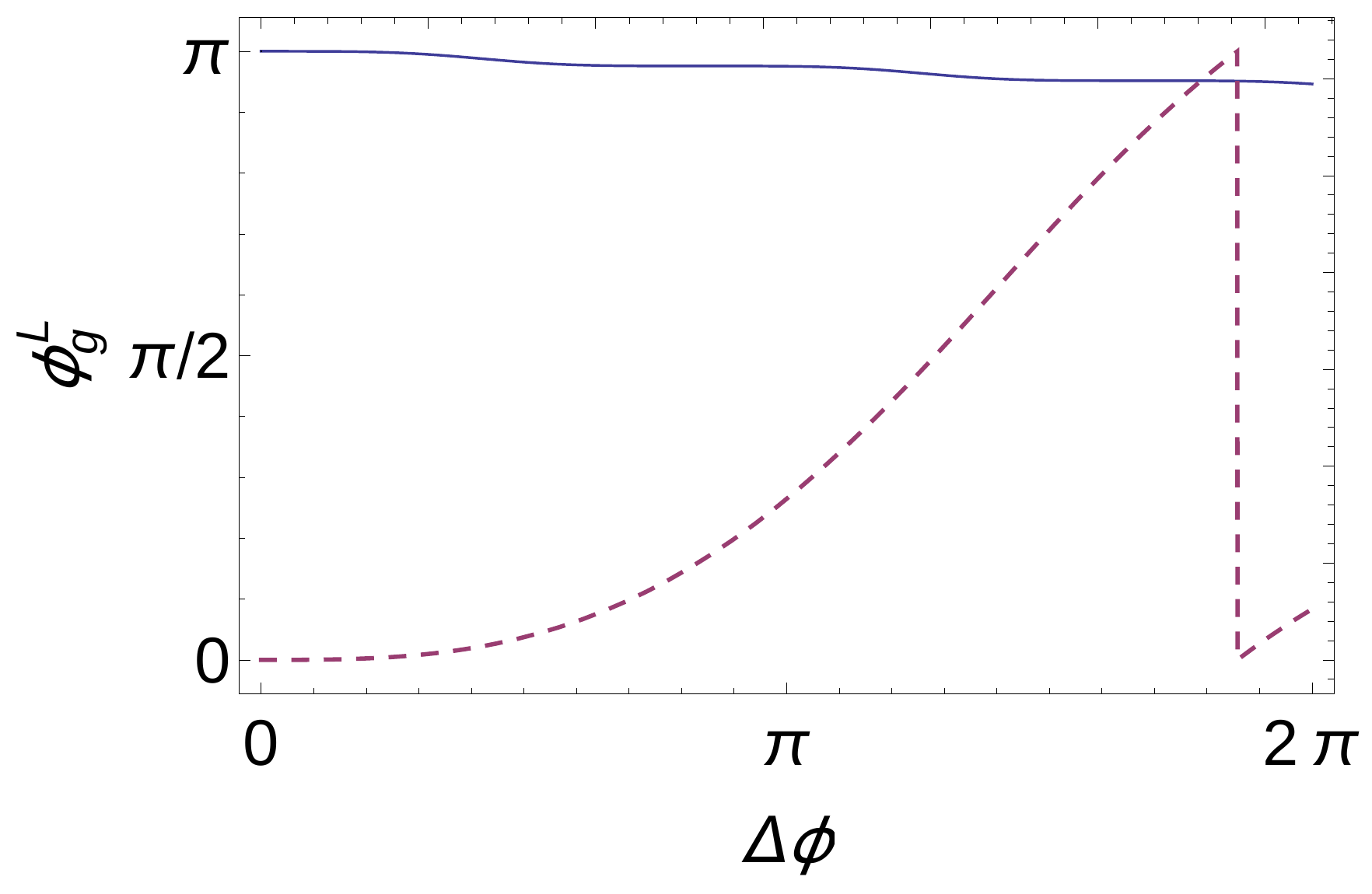}
  \caption{ Noncyclic geometric phase for the cases $\dot{\phi}= 100$ (solid curve) and $\dot{\phi}= -200$ (dotted curve)}
  \label{fig:gp(b)}
\end{subfigure}
\vspace{-2mm}
\caption{Geometric phases associated with the curves in the Bloch sphere for neutrino spin-precession $\nu_{eL}\rightarrow \nu_{eR}$.} 
 \label{fig:gp_ncyc}
\end{figure*}
The area enclosed by the trajectory traced out by neutrino spin rotation in projective Hilbert space, which in this case is  Bloch sphere $S^2$, is related to the geometric phases acquired by the neutrino state during the evolution. If a neutrino is initially created in the left-helicity state, i.e., $\Ket{\nu(0)}= (0 \quad  1)^T$, then after traveling a distance $z$ in the magnetic field the neutrino eigenstate will be a mixture of left- and right-handed components $\Ket{\nu(z)}= (\nu_R(z) \quad \nu_L(z))^T$ .  The geometric phase associated with the curve $\mathcal{C}$ traced  by the state $\Ket{\nu(z)}$ on the Bloch sphere is then given by
\vspace{-1mm}
\begin{equation}\label{gpl}
\phi_g^{L}[\mathcal{C}]= \arg \braket{\nu(0)|\nu(z)}- \Im \int_0^z \braket{\nu(z^\prime)|\frac{d}{dz^\prime}| \nu(z^\prime)} dz^\prime.
\end{equation}
Using Eq.\eqref{state_nuL}, we get the following expressions for the geometric phase: 
\begin{equation} \label{gp-left}
\begin{aligned}[b]
\phi_g^L[\mathcal{C}]=& -\arctan\Big(\cos 2 \theta_m \tan\frac{\phi_p}{2}\Big) + \frac{\phi_p}{2} \cos 2 \theta_m \\& +\frac{\Delta \phi}{2}\sin ^2 2\theta_m \Big(1-\frac{\sin \phi_p}{ \phi_p} \Big), 
\end{aligned}
\end{equation}
where $\Delta \phi= \phi(z)- \phi(0)$. Similarly if a neutrino is produced initially in the right-helicity state, the geometric phase acquired is  
\begin{equation} \label{muk-sim-r}
\begin{aligned}[b]
\phi_g^R =&- \phi_g^L. 
\end{aligned}
\end{equation}
Hence the spin and spin-flavor evolution of neutrino helicity states involve nonzero geometric phases. These expressions for geometric phases are valid regardless of whether the neutrino propagation is adiabatic or not, unlike the case of the Berry phase which  requires the propagation to be adiabatic.

Two particular cases clearly bring out the relation between the geometric phase and area enclosed by neutrino spin trajectory on the Bloch sphere. In the cyclic limit, as the neutrino spin-vector $\textbf{n}$ returns to its initial position i.e. $\phi_p = 2 \pi$, the geometric phase for each cycle is given by Eq.\eqref{gp-left} as
\begin{equation}\label{gp-cyc}
\phi_g^L[\mathcal{C}]\Big|_{cyc}=  -\pi(1- \cos 2 \theta_m)+ \frac{\Delta \phi}{2}\sin ^2 2\theta_m . 
\end{equation}
This result is particularly easy to visualize in the rotating frame of the magnetic field where $\Delta \phi = 0$. In this frame the geometric phase reduces to the famous value $-\pi(1- \cos 2 \theta_m)$, which is equal to $- \frac{1}{2}$ of the solid angle subtended, by the neutrino spin rotation trajectory on the Bloch sphere, at the center of the sphere. Another case is that of resonance condition \eqref{resonance}, for which Eq.\eqref{gp-cyc} gives the geometric phase: 
\begin{equation}
\phi_g^L[\mathcal{C}]\Big|_{\rm res} = 0. 
\end{equation}
This is expected since the resonance condition corresponds to the case when the neutrino trajectory traces out a great circle in the $x-z$ plane in the rotating frame. This is akin to parallel transport of a vector along a geodesic, which does not give rise to holonomy. The  corresponding curve in Fig. \ref{fig:bloch}(b) encloses no net oriented area, and thus has zero geometric phase.  

For the case of noncyclic evolutions the geometric phase can be interpreted in terms of a solid angle subtended by the neutrino spin rotation curve obtained by geodesic closure on the Bloch sphere. In Fig. \ref{fig:gp_ncyc}, we plot the variation of the geometric phase with the relative phase shift of the magnetic field for the case of neutrino spin precession $\nu_{eL}\rightarrow \nu_{eR}$.

Next we will study the neutrino spin and spin-flavor evolution in the case of a neutron star with realistic density and magnetic field profiles. We will examine various cases both inside and outside the neutron star and analyze the quantitative difference in geometric phases in different scenarios. 

\section{Neutrino Propagation in Neutron Stars}	
When stars run out of nuclear fuel at the end of their lives, the core of the star collapses under its own gravity resulting in a supernova explosion. Neutron stars (NSs) are the compact objects that are formed as final remnants of the core collapse supernova of stars with a mass of about $8-20$ times the mass of the Sun. NSs contain some of the most extreme astrophysical environments where the interior densities can be $\sim 5-10$ times the nuclear saturation density ($\approx 2.8 \times 10^{14}$ ${\rm g/cm}^3$) and where magnetic fields from the surface to the interiors can vary from $10^{15}$ to $10^{18}$ G \cite{Usov:1992zd,Lai:1991}. Although NSs are primarily composed of neutrons, there is also a small fraction of protons, electrons, and other nuclei. In the interior where density exceeds nuclear saturation density, exotic particles such as deconfined quarks, stable hyperon matter, and superfluid pion condensate may appear \cite{Baym:1975mf,Lattimer:2004pg}.

\begin{figure*}[ht]
\centering
\begin{subfigure}{.5\textwidth}
  \centering
  \includegraphics[width=.8\linewidth]{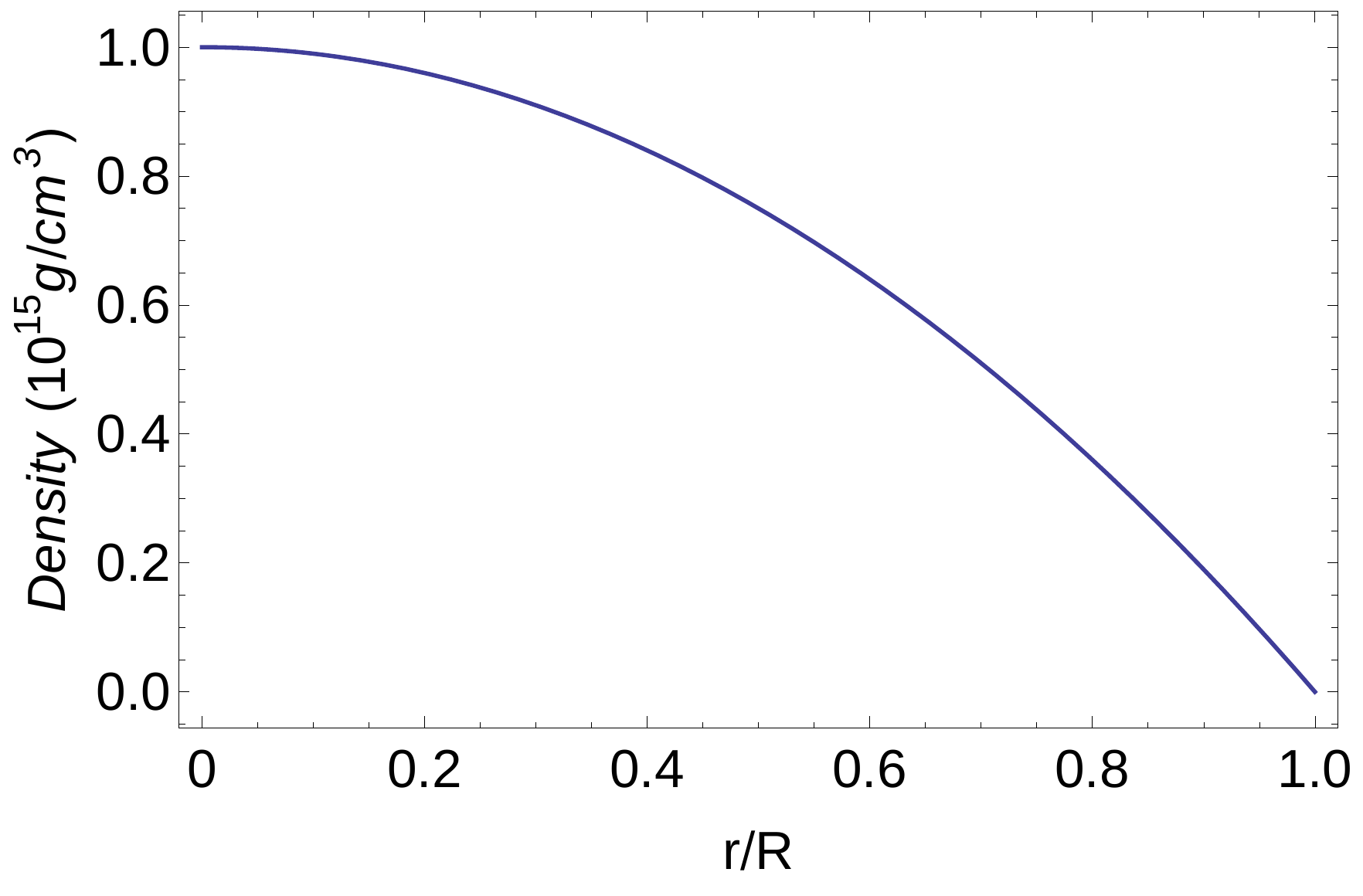}
  \caption{\centering}
  \label{fig:sub1}
\end{subfigure}%
\begin{subfigure}{.5\textwidth}
  \centering
  \includegraphics[width=.8\linewidth]{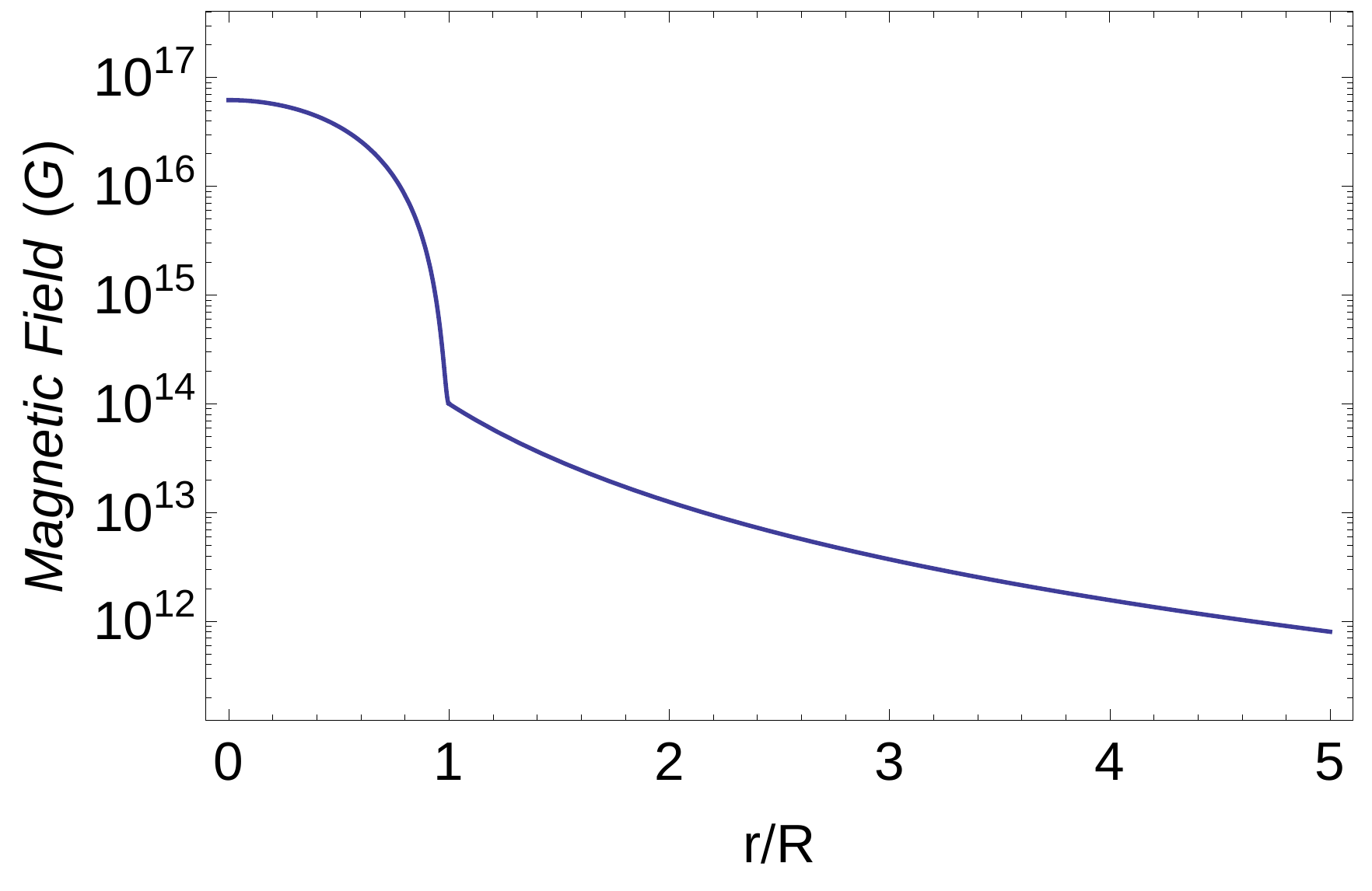}
  \caption{\centering}
  \label{fig:sub2}
\end{subfigure}
\caption{(a) Density and (b) magnetic field profiles of the neutron star. Magnetic field is plotted in log scale.}
\label{fig:profile}
\end{figure*}

Neutrinos play an important role in the formation and subsequent cooling of NSs. During the first few seconds of the supernova collapse a large number of neutrinos diffuse through the resulting proto-NS, which leads to a rapid drop in temperature by a factor of $\sim 100$. After about a minute, the NS becomes transparent to neutrinos resulting in a further drop in temperatures. The main process by which the neutrinos are produced in the NS cores is so-called direct Urca process $n\rightarrow p+ e^- + \bar{\nu}_e$, $p+e^-\rightarrow n+ \nu_e$\cite{Shapiro:1983du,Lattimer:2004pg}. However, the direct Urca process requires a certain energy threshold below which the neutrino emission occurs via modified Urca process $n+ (n,p) \rightarrow p+ (n,p)+ e^-+ \bar{\nu}_e$, $p+ (n,p)\rightarrow n+ (n,p)+ e^++ \nu_e$ \cite{Chiu:1964zza}. There are several other mechanisms by which neutrinos are produced in the NSs and help in the NS cooling (see \cite{Yakovlev:2000jp} for a detailed review).

In the following we study the spin-flavor evolution of the neutrinos produced in the core region of the NS. For definiteness we consider only the left-handed electron neutrinos produced below the resonance region and calculate the geometric phases they acquire as they propagate in the interior regions of the NS and finally come out of it. We also study the effect of magnetic field rotation on the geometric phases and the probabilities of the spin and spin-flavor conversion. These calculations require the knowledge of the density and magnetic field profiles in the interior and outer regions of the NS. The knowledge of the exact density profile of  NS depends strongly on the equation of state for which many models have been proposed (see \cite{Lattimer:2015nhk} for a recent review). However, without going into details of the models, we assume a simplistic density profile where the density decreases quadratically from the center
\begin{equation} \label{density}
\rho Y_e^{\textrm{eff}} = \begin{cases} \rho_0 + \rho_1 r^2, & \mbox{for } r \leq R \\ 
						 0  & \mbox{for } r > R \end{cases},
\end{equation}
where $R$ is the radius and $\rho_0$ is the central density of the NS. The values of radius and central density are taken as $ R= 10$ km and $\rho_0= 10^{15}$ ${\rm g /cm}^3$. The typical surface density of the NS is $\sim 10^{9}$ ${\rm g/cm}^3$ which determines the value of $\rho_1$.The magnetic field profile in the interior\cite{Bandyopadhyay:1997kh} and outer\cite{Likhachev:1990ki} regions of the NS are taken as 
\begin{equation} \label{mag_field}
B(r) = \begin{cases} B_s + B_c \bigg(1- \exp\Big(-\beta (\rho/\rho_s)^\gamma\Big)\bigg) & \mbox{for } r \leq R \\ 
						 B_s (R/r)^3 & \mbox{for } r > R \end{cases},
\end{equation}
where $ \beta= 0.005 $,  $ \gamma= 2 $,  $ B_c= 10^{18}$ G, $ B_s= 10^{14}$ G, and $\rho_s$ is the nuclear saturation density. The density and magnetic field profiles for the NS are plotted in Fig. \ref{fig:profile}.

\subsection{Geometric phases}
\begin{figure*}[tbph]
\centering
\begin{subfigure}{.5\textwidth}
  \centering
  \includegraphics[width=.8\linewidth]{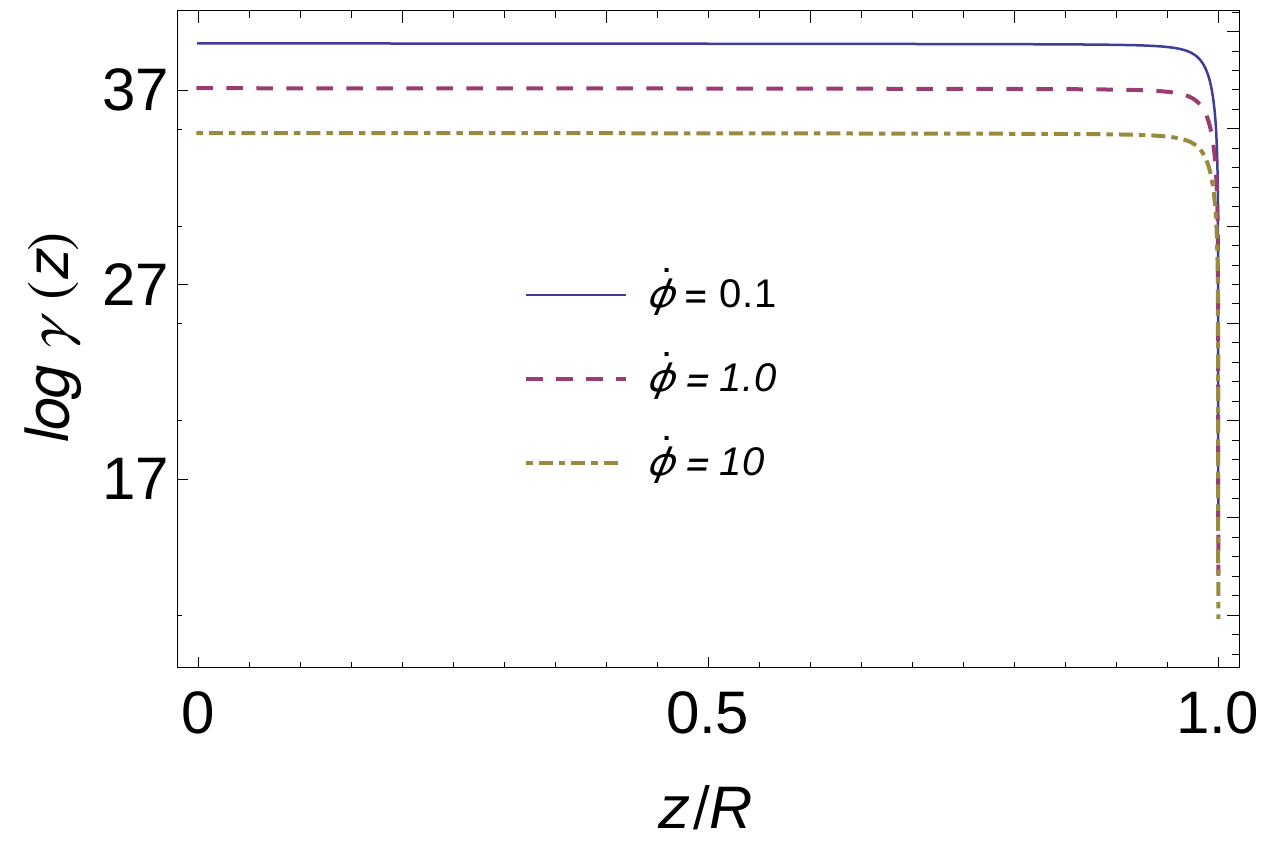}
  \caption{\centering}
\end{subfigure}%
\begin{subfigure}{.5\textwidth}
  \centering
  \includegraphics[width=.8\linewidth]{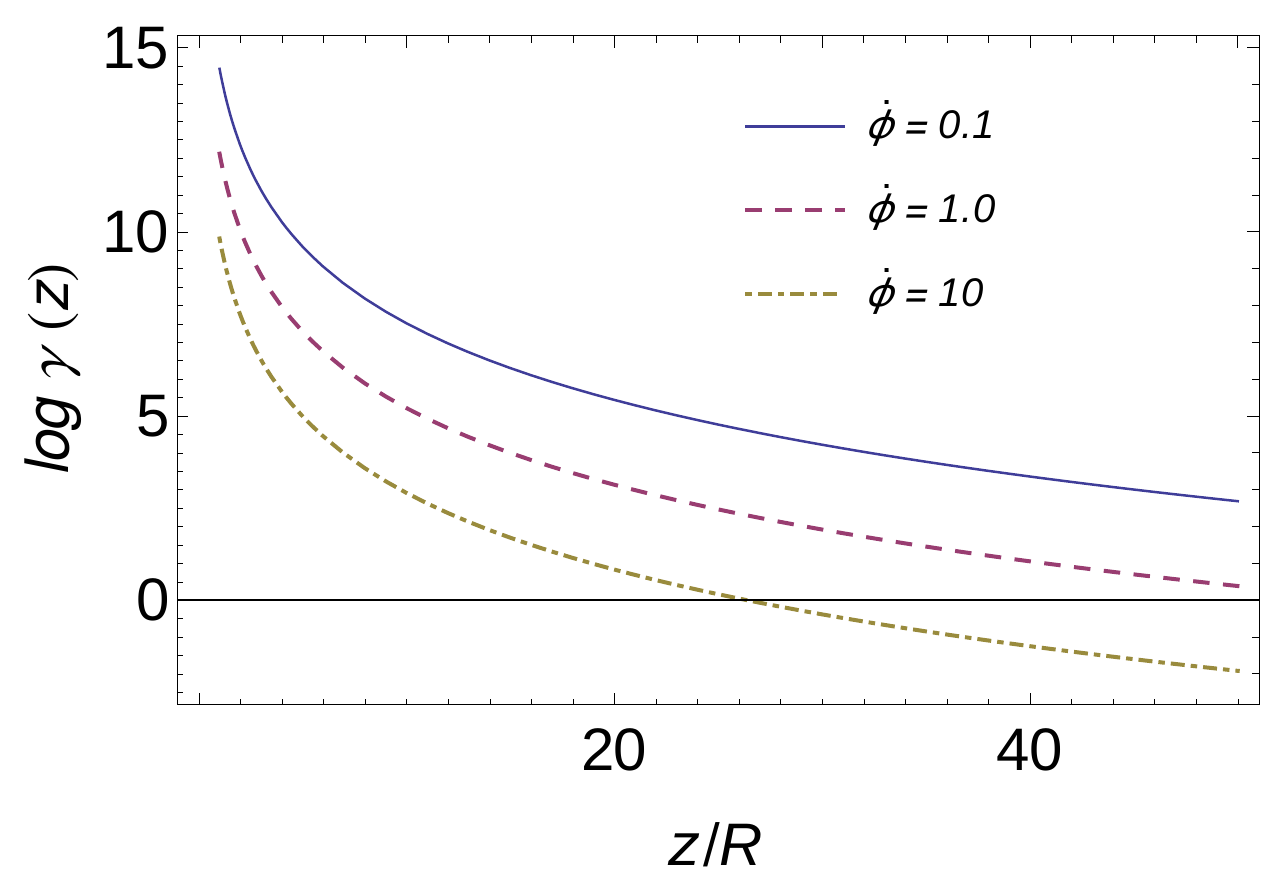}
  \caption{\centering}
\end{subfigure}
\caption{$\log \gamma(z)$ as a function of distance (a) inside and (b) outside the NS. The adiabaticity condition $\gamma \gg 1$ is satisfied for all values of $\dot{\phi}$ while for the outside regions adiabaticity holds only in a limited region.}
\label{fig:log_gamma}
\end{figure*}

The Hamiltonian for the neutrino spin-flavor evolution equation \eqref{evol2} can be written as
\begin{equation} \label{evol4}
H= -\frac{1}{2} \bm{\sigma} \cdot \textbf{B}_\textbf{eff}(z),
\end{equation}
where 
\begin{align}
\textbf{B}_\textbf{eff}=& |B_{\rm eff}|(\sin \chi(z) \cos \phi(z), \sin \chi(z) \sin \phi(z), \cos \chi(z)), \label{beff} \\
|B_{\rm eff}|=& \hspace{2mm} \sqrt[]{V(z)^2+ (2 \mu B(z))^2}, \\
\chi=& \tan^{-1}\bigg(\frac{2 \mu B(z)}{V(z)}\bigg), \\
V=& \frac{\sqrt[]{2} G_F \rho Y_e^{\rm eff}}{m_N}- \frac{\Delta m^2}{2E}\cos 2 \theta.
\end{align}
Formally, the solution of Eq.\eqref{evol1} with Hamiltonian \eqref{evol4} is given by the evolution matrix 
\begin{equation} \label{evol_in}
S(z,z_0)= \mathcal{P} \exp\bigg(-\frac{i}{2}\int_{z_0}^{z} (V(z') \sigma_3+ 2\mu B(z') \sigma_1) dz'  \bigg),
\end{equation}
where $\mathcal{P}$ is the path ordering operator.However, in the limit of adiabatic approximation the state of the system is given by one of the instantaneous eigenstates of the Hamiltonian \eqref{evol4}. The eigenstates, representing the spin polarization along and opposite to the direction of $\textbf{B}_\textbf{eff}$ are given by 
\begin{align}
\Ket{\psi_+}=& \begin{pmatrix}
\cos \chi(z)/2  \\ e^{i \phi(z)} \sin \chi(z)/2 
\end{pmatrix} \label{psiplus}, \\
\Ket{\psi_-}=& \begin{pmatrix}
e^{-i \phi(z)} \sin \chi(z)/2  \\ -\cos \chi(z)/2
\end{pmatrix} \label{psiminus},
\end{align}
corresponding to the eigenvalues $\mp |B_{\rm eff}|/2$. If the initial spin polarization of the neutrino is along the direction of magnetic field then the state of the system is represented by $\Ket{\psi_+}$, and the adiabatic condition is given by 
\begin{equation}\label{adia1}
\Big|\frac{\braket{\psi_{+}| \dot{\psi}_-}}{E_{+}- E_-}\Big| \ll 1,
\end{equation}
which is equivalent to  $\sqrt[]{\dot{\chi}^2+ (\dot{\phi} \sin \chi)^2}/2 \ll |B_{\rm eff}|$.
We define the adiabaticity parameter 
\begin{equation}
\gamma= \frac{|B_{\rm eff}|}{\sqrt[]{\dot{\chi}^2+ (\dot{\phi} \sin \chi)^2}},
\end{equation}
so the adiabaticity condition \eqref{adia1} is equivalent to $\gamma \gg 1$. We now calculate $\gamma$ for regions both inside and outside the NS. We find that, while in the inside region the adiabaticity holds for practically all values of $\dot{\phi}$, in the outside region of NS the range over which the adiabatic solution is valid is restricted, depending on the values of $\dot{\phi}$.  The larger the value of $\dot{\phi}$, the smaller is the region over which the adiabatic approximation is valid. For typical values of $\dot{\phi}$, the range over which adiabaticity holds is roughly $20 - 30$ times the radius of the NS as shown in Fig. \ref{fig:log_gamma}.

In this case the magnetic field \eqref{beff} traces out an open curve $C_R$ in the parameter space $R^3$. Under adiabatic evolution the noncyclic geometric phase associated with the curve $C_R$, for the case of neutrino with initial spin polarization along the direction of magnetic field, is given by the generalization of Berry's phase \cite{GarciadePolavieja:1998vy} 
\begin{equation}
\begin{aligned}[b]\label{gp_adia}
\phi_g[C_R]=& {\rm arg} \Braket{\psi_+(0)| \psi_+ (z)} \\ &- \Im \int_0^z d\mathbf{R} \cdot \Braket{\psi_+(\mathbf{R})|\mathbf{\nabla}_R|{\psi_+(\mathbf{R})}},
\end{aligned}
\end{equation}
where $\textbf{R}$ represents the magnetic field \eqref{beff}. Using Eq.\eqref{psiplus} we calculate the geometric phase as 
\begin{widetext}
\begin{equation} \label{gp_ns}
\phi_g^+[C_R]= \tan^{-1}\bigg(\frac{\sin \Delta \phi(z) \sin \frac{\chi(z)}{2} \sin \frac{\chi(0)}{2}}{\cos \frac{\chi(z)}{2} \cos \frac{\chi(0)}{2}+\cos \Delta \phi(z)\sin \frac{\chi(z)}{2} \sin \frac{\chi(0)}{2}}  \bigg) -\frac{\Delta \phi(z)}{2}(1- \cos \chi(z)).
\end{equation}
\end{widetext}
While for the other eigenstate the geometric phase is $\phi_g^-[C_R]= - \phi_g^+[C_R]$. 

\begin{figure*}[tbph]
\centering
\begin{subfigure}{.5\textwidth}
  \centering
  \includegraphics[width=.8\linewidth]{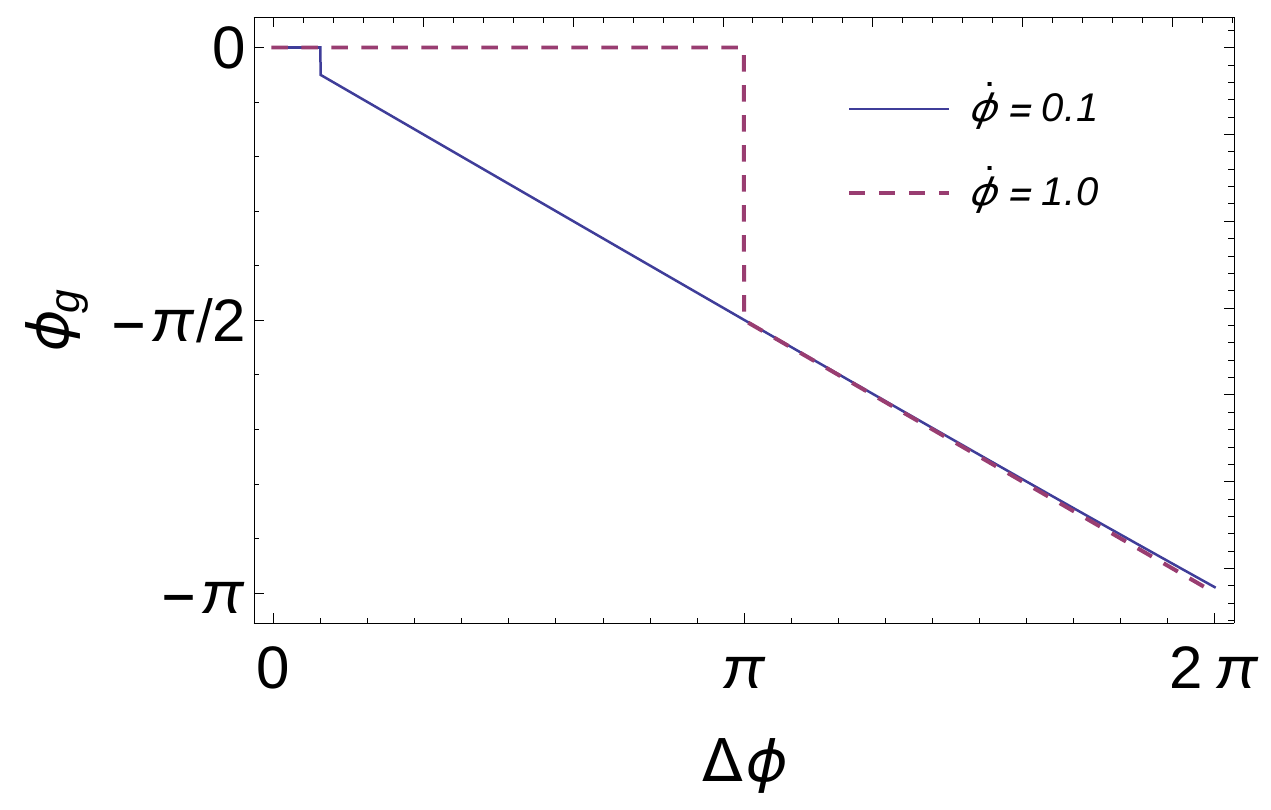}
  \caption{\centering}
\end{subfigure}%
\begin{subfigure}{.5\textwidth}
  \centering
  \includegraphics[width=.8\linewidth]{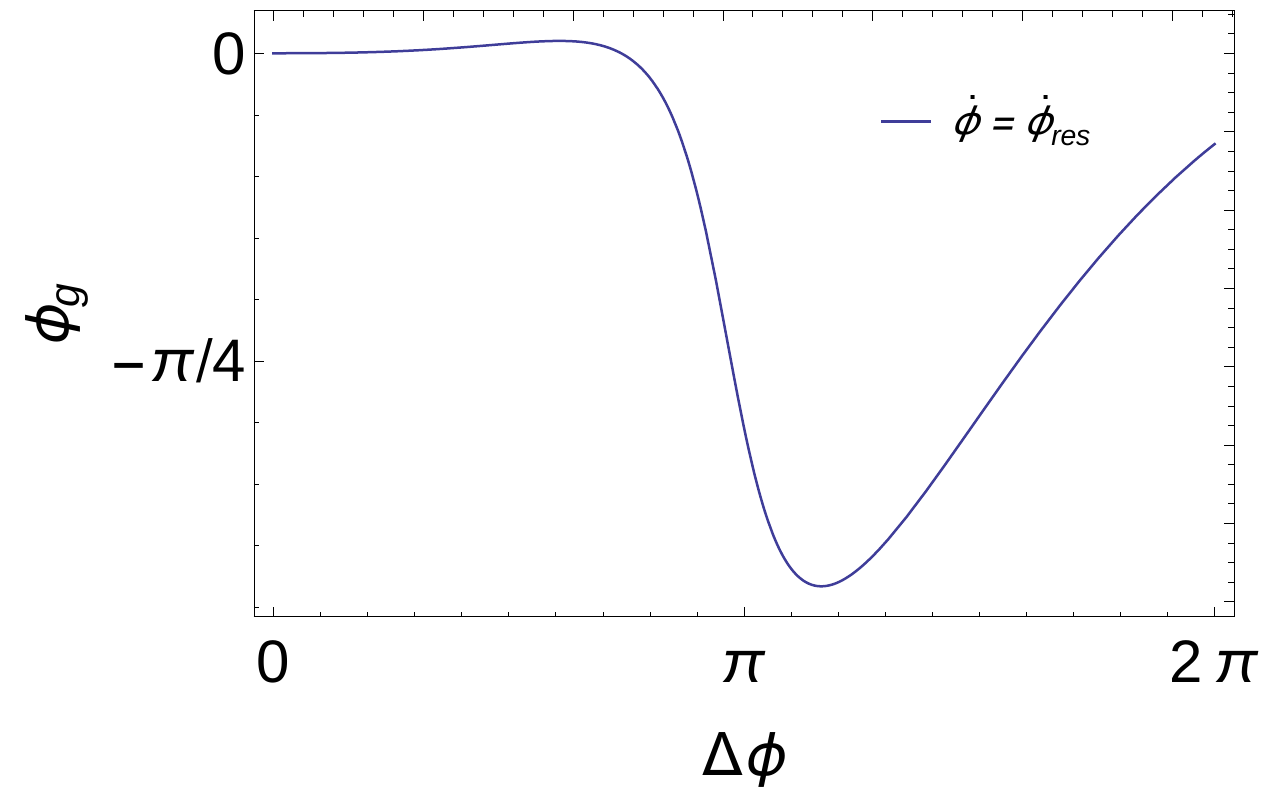}
  \caption{\centering}
\end{subfigure}
\caption{\raggedright{Geometric phases neutrino propagation in NS. In (a) the flat portion of the curve corresponds to neutrino propagation inside the NS, where the geometric phase is almost zero. In (b) $\dot{\phi}_{\rm res}$ corresponds to the resonant condition $V= -\dot{\phi}$.    }}
\label{fig:gp_adia_ns}
\end{figure*}

Since the definition \eqref{gp_adia} assumes adiabaticity, the expression \eqref{gp_ns} is valid only when the adiabatic condition \eqref{adia1} is satisfied.  When the nonadiabatic effects arise, one has to resort to more general methods such as that of geodesic closure to calculate geometric phases. However, we are only interested in the qualitative features of the  geometric phases that arise due to neutrino spin and spin-flavor oscillations in the NS environment. Since in the inside region of the NS, the matter effects strongly dominate over the magnetic field, the area of the  curve traced by $\textbf{B}_\textbf{eff}$ is negligible, and hence the associated geometric phase is vanishingly small. As the neutrinos come out of the NS, matter effects vanish and now neutrino eigenstates develop a significant geometric phase as shown in Fig. \ref{fig:gp_adia_ns}.  

\subsection{Transition probabilities and cross boundary effect}
\begin{figure*}[htp]
\begin{subfigure}{.5\textwidth}
  \includegraphics[width=0.9\linewidth]{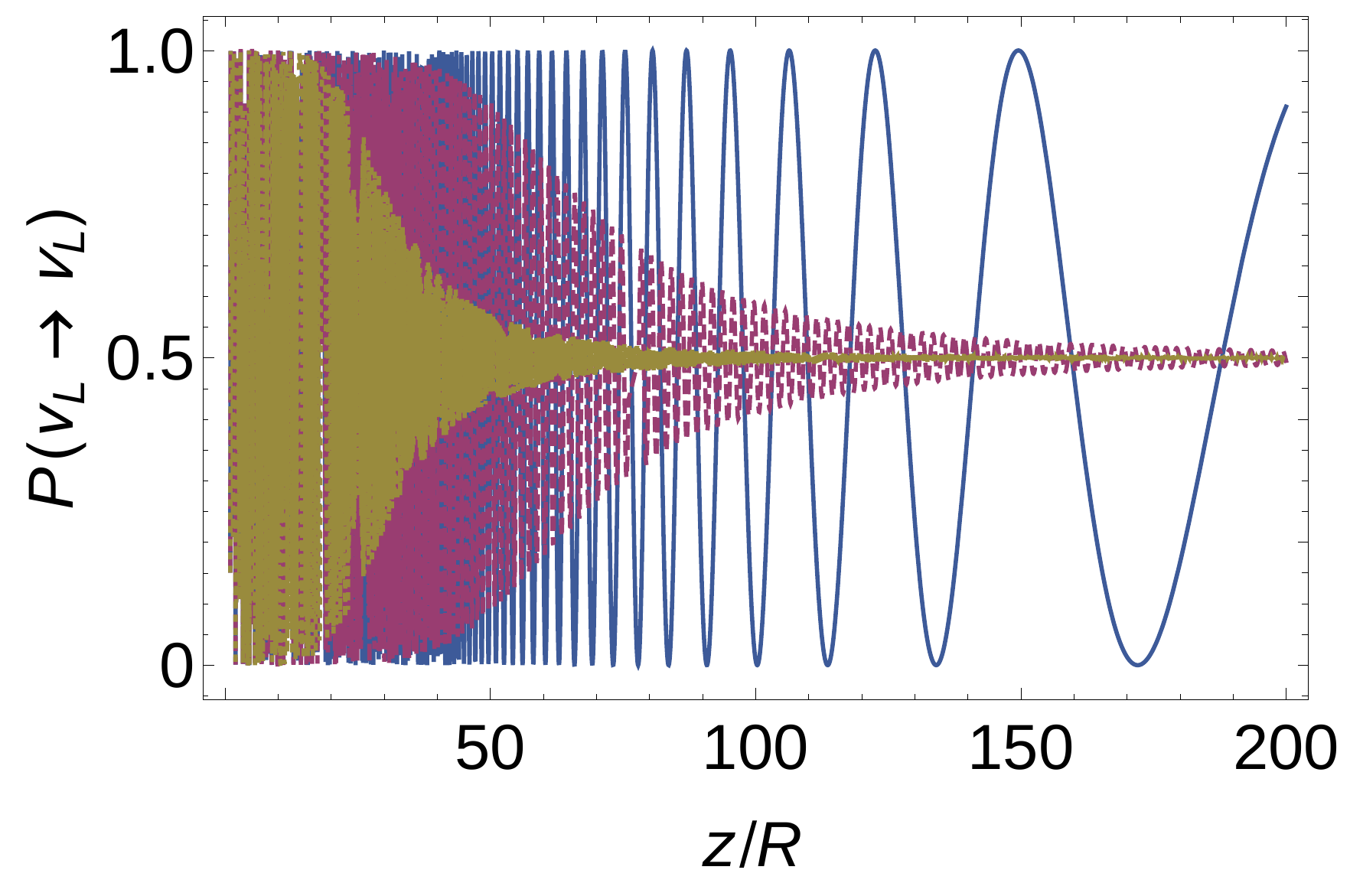}
  \caption{\centering Spin transition probability, $\nu_{eL} \rightarrow \nu_{eR}$ }
\end{subfigure}%
\begin{subfigure}{.5\textwidth}
  \includegraphics[width=.9\linewidth]{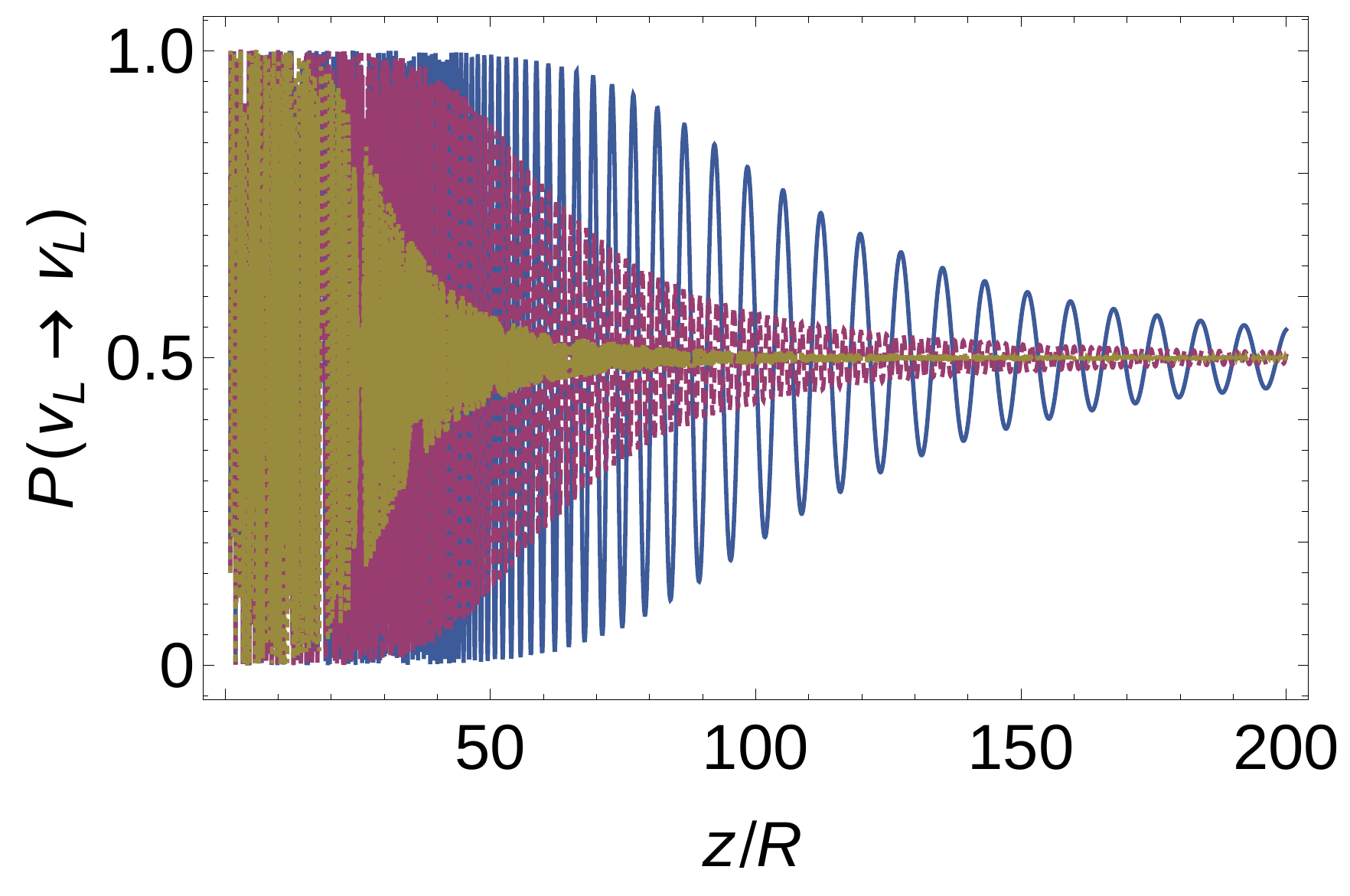}
   \caption{\centering Spin-flavor transition probability,  $\nu_{eL} \rightarrow \nu_{\mu R}, \bar{\nu}_\mu$ }
\end{subfigure}
\begin{subfigure}{\linewidth}
\centering
 \includegraphics[width=0.3\textwidth]{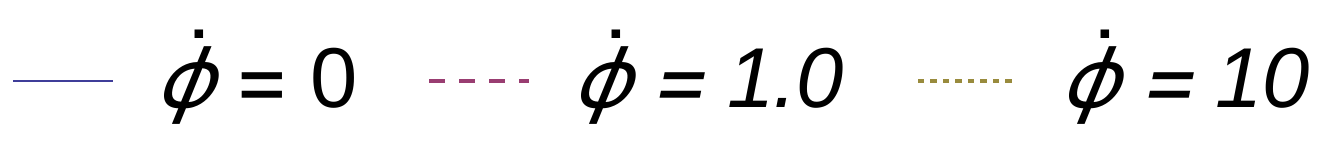}  
\end{subfigure}
\vspace{-2mm}
\caption{  Neutrino survival probability of spin and spin-flavor precession for various values of the rotation frequency. Nonzero values of $\dot{\phi}$ lead to suppression of  transitions and the probability converges to one-half at a faster rate compared to the case when $\dot{\phi}= 0$. For the case of spin transitions in nonrotating magnetic fields the probability does not converge to $0.5$ but instead approaches 1 in the limit $z \gg R$. This is because for this case $ \cos \theta_{\rm eff}= 0$ and the oscillatory term in Eq.\eqref{prob_ns} converges to 1 in the limit $z\gg R$.}
\label{fig:prob_ns}
\end{figure*}

Now we calculate the spin and spin-flavor transition probabilities as the neutrinos propagate in NS's under adiabatic conditions. Considering the case of left-handed electron neutrinos produced near the center of the NS, the adiabatic survival probability is given by \cite{Giunti:2007ry,Likhachev:1990ki}
\begin{equation} \label{prob_ns}
\begin{aligned}[b]
P(\nu_L \rightarrow \nu_L)(z)=& \frac{1}{2}\big(1+\cos \theta_{\rm eff}(z_0) \cos \theta_{\rm eff} (z)\\& + \sin \theta_{\rm eff}(z_0) \sin \theta_{\rm eff}(z) \cos \zeta(z)\big),
\end{aligned}
\end{equation}
where 
\begin{align}
\theta_{\rm eff}(z)=& \tan^{-1} \Big(\frac{2 \mu B}{V+ \dot{\phi}}\Big),\\
\zeta(z)=& \int_{z_0}^z dz' \hspace{2mm} \sqrt[]{(V+ \dot{\phi})^2+(2 \mu B)^2}.
\end{align}
For the neutrino propagation inside the NS, for the given density and magnetic field profile $V \gg 2 \mu B$, and hence $\theta_{\rm eff} \approx 0$, so according to Eq.\eqref{prob_ns} $P(\nu_L \rightarrow \nu_L) \approx 1$. Thus there are almost no spin or spin-flavor transitions inside the NS. However, for the outside case the situation is more interesting and there are appreciable transitions as shown in Fig. \ref{fig:prob_ns}. After about $200 R$ half of the left-handed neutrinos produced inside the NS are converted into the right-handed neutrinos.

For the case of neutrino propagation in a medium of constant density and uniformly twisting magnetic fields, one can define a critical magnetic field, which is the magnetic field required for the oscillation amplitudes $\nu_L \rightarrow \nu_R$ to be close to unity and is given by \cite{Likhachev:1990ki}
\begin{equation} \label{Bcr}
\begin{aligned}[b]
B_{cr}[G]=& 43 \bigg(\frac{\mu_B}{\mu}\bigg)\bigg| \bigg(\frac{\Delta m^2}{1 eV^2}\bigg)\bigg(\frac{MeV}{E}\bigg)  \cos 2 \theta \\& - 2.5 \times 10^{-31}\bigg(\frac{n_{\rm eff}}{{\rm cm}^{-3}}\bigg)+ 0.4\bigg(\frac{\dot{\phi}}{m} \bigg) \bigg|.
\end{aligned}
\end{equation}
By calculating $B_{cr}$ for different situations, we can get rough estimates of the magnetic field required for appreciable neutrino transitions. 
In the interior regions of the neutron star, $\rho \approx 10^{15}$ ${\rm g/cm}^3$ gives 
$n_{\rm eff} \approx 6 \times 10^{38}$ ${\rm cm}^{-3}$, hence for neutrinos with energy $E= 1$ MeV, Eq.\eqref{Bcr} gives $B_{cr} \approx 6 \times 10^{20}$ G. Since $B_{cr} \gg B $ in the interior of the neutron star, the transitions are negligible, as also shown by the probability argument above. Even in the 
outermost crust of the star,  $n_{\textrm{eff}} \approx 10^{33}$ ${\rm cm}^{-3}$, and $B_{cr} \approx 10^{15} G$, which is greater than the magnetic fields prevailing in those regions. So we expect very weak neutrino transitions inside the neutron stars in the case of neutrinos produced below the resonance regions. For the regions just outside the neutron star density suddenly drops to zero, so there is a sharp decrease in the critical magnetic field required for the helicity transitions. For $1$ MeV neutrinos, Eq.\eqref{Bcr} gives $B_{cr} = 10^8$ G. Since the magnetic field just outside the neutron star is $\sim 10^{14} G(\gg B_{cr})$, as the neutrinos cross the surface of the NS, there are rapid helicity transitions that are termed as cross boundary effects\cite{Likhachev:1990ki}. 

Since the magnetic field outside the NS falls off as $1/r^3$, the range over which the magnetic field exceeds critical magnetic field is given by $r_{cr}= R (B/B_{cr})^{1/3} \approx 100 R$. As can be seen in Fig. \ref{fig:prob_ns}, the oscillation amplitude reduces as we go away from the NS and almost vanishes for $r> 200 R$ in the case for nonrotating fields. If we consider the effect of field rotation then according to Eq.\eqref{Bcr} the critical magnetic field required to sustain oscillations increases. For $\dot{\phi}= 10$, the $B_{cr} \approx 6 \times 10 ^8$ G. Thus the range over which oscillation amplitudes are finite decreases to $\approx 50 R$.

\section{Possible methods of geometric phase detection}
The usual method of geometric phase detection employs experiments wherein a beam is split into two parts, both parts undergo evolution along different paths in parameter space, and then they are made to interfere. The resulting interference pattern bears the signature of the geometric phase. However, these types of experiments are not feasible in case of neutrinos due to their small interaction cross section that renders them practically impossible to maneuver.

Another approach, which has become popular in recent years, is that of quantum simulation. In this approach quantum systems that cannot be accessed experimentally are simulated using a controllable physical system underlying the same mathematical model \cite{Georgescu:2013oza}. The possibility of studying neutrino systems by quantum simulation has been explored in \cite{Wang:2015ams,Wang:2015tqp}. In \cite{Wang:2015tqp} it was proposed to detect the neutrino geometric phases using the nuclear magnetic resonance (NMR) setup with a controllable range of parameters. Here we propose an analogous NMR experiment where parameters can be varied to simulate the environment of neutrino oscillations in magnetic fields. The Hamiltonian for a standard NMR experiment is given by \cite{suter1987berry}
\begin{equation} \label{nmr-hamiltonian}
H= -\frac{\omega_0}{2}  \big[\cos \theta \sigma_z+ \sin \theta(\sigma_x \cos \omega t+ \sigma_y \sin \omega t) \big],
\end{equation}
where $\omega_0$ is the Larmor precession frequency of the spins, $2 \theta$ is the angle between the magnetic field direction and the quantization axis, and $\omega$ is the frequency of the circularly polarized magnetic field. 
Comparing Eq.\eqref{nmr-hamiltonian} with the Hamiltonian \eqref{evol2} for neutrinos we get the following values for the NMR parameters:
\begin{align}
\omega_0=& \hspace{2 mm} \sqrt[]{V^2+ (2 \mu B)^2},\\
\theta = & \tan^{-1}\Big(\frac{2 \mu B}{\sqrt[]{V^2+ (2 \mu B)^2}}\Big),\\
\omega = & \dot{\phi}.
\end{align}
For example, the neutrino oscillation environment outside the NS can be simulated using the following range of parameters: $\omega_0/2 \pi \in (10^6- 10^3)$ MHz, $\theta \approx \pi/5$, and $\omega/2 \pi \in (1.5- 150)$ kHz. In this way, the geometric phases that arise in neutrino systems can be inferred from those obtained in NMR experiments with a suitably chosen range of parameters.

\section{Conclusions}
In this work, we have studied the noncyclic geometric phases associated with neutrino spin and spin-flavor transitions. The dynamics of neutrino spin rotation was examined in the Bloch sphere representation, which clearly brings out the geometric nature of this phenomena. The geometric phase acquired by a  neutrino state was shown to be related to the area enclosed by the curve traced by a neutrino spin vector. For the case of cyclic evolution, it was shown that the expressions reduce to the usual Aharonov-Anandan phase for spin precession in a magnetic field.  As a particular case, we analyzed the geometric phases acquired by the solar neutrinos as they propagate outwards under the effect of matter and magnetic fields of the Sun. 

Further, we analyzed the situation of neutrinos produced in the NS, propagating outwards under the effect of matter and magnetic fields. We have obtained analytical expressions of the noncyclic geometric phases in the adiabatic approximation and studied their behavior both inside and outside the NS for various cases. We have also studied the transition probability and the cross boundary effects and showed that at a distance of about 200 times the radius of a NS, the initial flux of left-handed neutrinos produced inside the NS is depleted to half of its original value. We would  like to point out that we considered only the case of neutrinos produced below resonance regions. However, there might arise situations where there may be significant resonant effects due to both matter and magnetic fields, and it would be interesting to explore these effects in the context of geometric phases.  

The emergence of geometric phases in neutrino spin and spin-flavor evolution highlights an important geometric aspect of this phenomena. Even though at present there seems to be no method to detect such phases directly in the current experiments, alternative methods such as quantum simulation have been proposed to detect such phases.  The present calculations bring out an essentially geometric character manifest in the neutrino spin rotation and is well worth exploring further. 

\section*{Acknowledgments}
The authors are thankful to the referee for critical comments and suggestions.

\bibliography{nuref,gpref,neutron_star}

\end{document}